\documentclass[12pt]{article}
\usepackage{amsfonts}
\usepackage{graphicx}
\usepackage{amsthm}
\usepackage{latexsym,amsmath,color}
\usepackage[round]{natbib}
\usepackage[colorlinks=true,linkcolor=black,urlcolor=black]{hyperref}
\usepackage[english]{babel}
\usepackage{multicol}
\usepackage{epstopdf}
\usepackage{subfigure}
\usepackage{float}
\usepackage[page]{appendix}
\usepackage{geometry}
\hypersetup{citecolor=black}

\usepackage{array}
\usepackage{tabulary}
\newcolumntype{K}[1]{>{\centering\arraybackslash}p{#1}}

\begin{document}

\title{Bootstrapping kernel intensity estimation for nonhomogeneous point processes depending on spatial covariates}
\author{M.I. Borrajo, W. Gonz\'alez-Manteiga and M.D. Mart\'inez-Miranda}
\date{}

\newtheorem{theorem}{Theorem}[section]
\newtheorem{conjecture}[theorem]{Conjecture}
\newtheorem{corollary}[theorem]{Corollary}
\newtheorem{lemma}[theorem]{Lemma}
\newtheorem{proposition}[theorem]{Proposition}

\newtheorem*{remark}{Remark}

\newtheorem{definition}{Definition}

\maketitle

\begin{abstract}
	In the spatial point process context, kernel intensity estimation has been mainly restricted to exploratory analysis due to its lack of consistency. Different methods have been analysed to overcome this problem, and the inclusion of covariates resulted to be one possible solution. In this paper we focus on de\-fi\-ning a theoretical framework to derive a consistent kernel intensity estimator using covariates, as well as a consistent smooth bootstrap procedure. We define two new data-driven bandwidth selectors specifically designed for our estimator: a rule-of-thumb and a plug-in bandwidth based on our consistent bootstrap method. A simulation study is accomplished to understand the performance of our proposals in finite samples. Finally, we describe an application to a real data set consisting of the wildfires in Canada during June 2015, using meteorological information as covariates.
\end{abstract}

\section{Introduction}
Point processes are a branch of spatial statistics whose main aim is to study the geometrical structure of patterns formed by objects (called events) that are distributed randomly in number and space. This type of data arise in many different fields such as ecology, \cite{Illian2009} and \cite{Law2009}; epidemiology, \cite{Diggle1990cancer} and  \cite{Gatrell1996}; astronomy, \cite{Astrostatistics} and \cite{Kerscher2000};  forestry, \cite{Stoyan2000}; seismology, \cite{Ogata1988}, \cite{Ogata1998}, \cite{OgataZhuang2006} and \cite{Schoenberg2011}. General theory on point processes as well as some classical applications, can be found in \cite{DaleyVereJones1988}, \cite{Moller2003}, \cite{Illianbook}, \cite{Diggle2013} and \cite{Baddeley2015}.

Modelling the first-order intensity function is one of the main aims in point process theory. Assuming a parametric model for the intensity function may be a way of estimating it, using for instance a likelihood score such as the Akaike Information Criteria (AIC), see \cite{Vanlis2000}, \cite{Moller2003} and \cite{Diggle2013} or pseudolikelihood procedures, see \cite{Waagepetersen2007}. In the Bayesian context \cite{Illian2012} proposed some models based on log-Gaussian Cox processes. However, these techniques can provide inappropriate estimates when the assumed model does not fit the real intensity function. Hence, there is an alternative through nonparametric methods such as quadrat counts and kernel estimation. 

\cite{Diggle1985} proposed the first kernel intensity estimator, based on the structure of the common kernel density estimator defined by \cite{Parzen1962} and \cite{Rosenblatt1956}, with the inclusion of an edge correction term. The main drawback of Diggle's proposal is its lack of consistency, which has almost li\-mi\-ted its use to exploratory analysis. To overcome this problem, \cite{CucalaThesis} developed asymptotic theory for Diggle's estimator, introducing the concept of ``density of events locations'' which is based on the idea that the intensity and the density functions differ only in a constant. 

The use of nonparametric methods implies to choose a bandwidth value, which determines the degree of smoothness to be considered in the estimation. The choice of the bandwidth parameter is crucial and it has motivated several papers in the literature in the recent decades, see for example \cite{Marron1988}, \cite{Scott1992} and \cite{Silverman1986} for an earlier full description of the problem. There has been great theoretical developments on this problem in areas of statistics such as density estimation and regression, meanwhile in the context of point processes it has received less attention. \cite{Diggle1985} proposed a bandwidth selector based on the minimisation of the mean squared error (MSE) of his estimator. Later, \cite{DiggleMarron1988} showed the equivalence, for Cox processes in the real line, between that procedure and  the standard least-squares cross-validation method used in kernel density estimation. This is an example of the strong connection between this two problems of density and intensity estimation. \cite{Brooks1991} proved the optimality of the least-squares cross-validation bandwidth for one-dimensional nonhomogeneous Poisson point processes. \cite{Isa2015} develop an extension of Cucala's theory to the two-dimensional case, and propose a two-dimensional bandwidth selection method based on bootstrap.

Marks and covariates are two different ways of including some extra information in a point process models, \cite[Chap. 5]{Illianbook}. The main difference between them is that marks are directly linked to the events, while covariates include information about the whole observation region. This second scenario is what we consider in this paper. \cite{Guan2008} develop a kernel intensity estimator, assuming that the intensity depends on some observed covariates through a continuous unknown function. This estimator turns out to be consistent under some hypotheses concerning the kernel, the boundary of the region and the pair correlation function. The author also deals with the problem of high-dimensional covariates proposing a method, based on sliced inverse regression, to reduce the number of them before applying the kernel techniques. \cite{Baddeley2012} also use information coming from covariates but in a slightly different way. They assume that the intensity depends on a continuous covariate and propose some intensity estimators based on local likelihood and kernel techniques. In the latest, the bandwidth parameter is chosen using the common rule-of-thumb for density estimation, \cite{Silverman1986}, directly applied to the point process pattern. 

There is an important methodological gap in point process with covariates that we are trying to fill. Our idea is to exploit the relationship between the density and the intensity functions and apply nonparametric techniques, frequently used in the density context, to improve the existing estimates for the first-order intensity function. Indeed, we want to provid point processes with covariates, particularly the first-order intensity function, with a well defined framework to guarantee the consistency of the proposed estimates and all the required tools, such as data-driven bandwidth selection methods, to be able to apply them in practice.

This paper is organised as follows. In Section 2, we make a brief overview on the existing methods in kernel intensity estimation. Section 3 is devoted to set up the new framework for kernel intensity estimation with covariates and to develop asymptotic theory for it. In Section 4 we propose a new smooth bootstrap method and we prove its consistency. Section 5 includes two new data-driven bandwidth selection methods: a rule-of-thumb based on assuming normality and a bootstrap bandwidth selector. An extensive simulation study is carried out in Section 6 to analyse the our new proposals and to compare them with the existing competitors. In Section 7, we apply all these methods to a real data set of wildfires in Canada. We finally draw some conclusions in Section 8.

\section{Kernel intensity estimation}
Let $X$ be a point process defined in a region $W\subset\mathbb{R}^2$, where $W$ is assumed to have finite positive area. Let $X_1, \ldots, X_N$ be a realisation of the process where $N$ is the random variable counting the number of events. The first-order intensity, from now on referred as intensity, is defined following \cite{Diggle2013} as: 
\begin{equation*}
\lambda(x)=\lim_{|dx|\to 0}\frac{E[N(dx)]}{|dx|}, 
\end{equation*}
where $|dx|$ denotes the area of an infinitesimal region containing the point $x\in\mathbb{R}^2$.

There is an extensive literature on parametric point process models and intensity estimation in this case, see \cite{Schoenberg2005}. However it is well known that we can obtain unreliable estimates when the assumed parametrisation deviates from the true intensity. This is the main reason that supports the use of nonparametric techniques. \cite{Diggle1985} proposed the first kernel intensity estimator for one-dimensional point processes, which has been easily extended to the plane:
\begin{equation*}
\hat{\lambda}^D_H(x)=\frac{\sum_{i=1}^NK_H\left(x-X_i\right)}{p_H(x)}, \quad x \in \mathbb{R}^2
\end{equation*}
where $H$ is a bandwidth matrix, $K$ denotes a kernel function, $K_H(x)=|H|^{-1/2}K\left(H^{-1/2}x\right)$ and $p_H=\int_W{|H|^{-1/2}}K(H^{-1/2}(x-y))dy$ is an edge correction term.

This estimator has been widely used during decades for exploratory a\-na\-ly\-sis, but the inference performed with it has been limited due to its lack of consistency. To overcome this problem, \cite{CucalaThesis} defined the ``density of events locations'' as $\lambda_0(x)=\lambda(x)/m$, where $m=\int_W{\lambda(x)dx}$ is the expected number of events lying on W. He proposes a kernel estimator:
\begin{equation*}
\hat{\lambda}_{0,h}(x)=\frac{1}{N}\sum_{i=1}^N \frac{1}{h}K\left(\frac{x-X_i}{h}\right)1_{\{N\neq0\}}, x\in\mathbb{R}
\end{equation*}
with $1_{\{\}}$ denoting the indicator function and $h$ a one-dimensional bandwidth parameter. He proves its consistency under an infill structure asymptotic framework. \cite{Isa2015} extended these ideas to the two-dimensional situation using bandwidth matrices, as it has been done in the context of multivariate density estimation.

Now, let $Z:W\subset \mathbb{R}^2 \rightarrow \mathbb{R}$ be a spatial continuous covariate that is exactly known in every point of the region of interest $W$ and $Z_1,\ldots, Z_N$ the realisation of the transformed process, i.e, $Z_i=Z(X_i)$. In practice, following the indications of \cite{Baddeley2012}, this covariate will commonly be know in an enough amount of points spread over the region, so the values for the rest of the points can be interpolated and it can be assumed that these values are indeed the real ones.

Once it is assumed that the spatial point process intensity depends only on the covariate, see \cite{Baddeley2012}, the following can be established
\begin{equation}\label{eq:int_rho}
\lambda(u)=\rho(Z(u)), \: u \in W\subset \mathbb{R}^2,
\end{equation}
where $\rho$ is an unknown function. As $Z$ is known, only $\rho$ needs to be estimated in order to obtain an estimate of $\lambda$, which is the target.

To this purpose it is necessary to deal with the transformed univariate point process, $Z(X)$, and establish the theoretical relationship between this one and the original spatial point process $X$. If $X$ is a Poisson point process in $W\subset \mathbb{R}^2$ with intensity function \eqref{eq:int_rho}, then $Z(X)$ is a Poisson point process in $\mathbb{R}$ with intensity $\rho g^\star$ and with the same expected number of events, where $g^\star$ is the non-normalised version of the derivative of the spatial cumulative distribution function (see Appendix A for details on this). 

The proposals in \cite{Guan2008} and \cite{Baddeley2012}, following assumption \eqref{eq:int_rho}, are similar kernel intensity estimators. \cite{Guan2008} develops a kernel estimator based on the definition of the distance between two points by the distance through their covariates values:

\begin{equation*}
\hat{\lambda}^G_{h}(u)=\frac{\sum_{i=1}^N K_h(||\mathbf{Z}(u)-\mathbf{Z}(X_i)||)}{q_h(u)},
\end{equation*}
with $q_h(u)=\int_W{K_h(||\mathbf{Z}(u)-\mathbf{Z}(s)||)ds}$ the edge correction term, where \linebreak $\mathbf{Z}=(Z_1,\ldots,Z_p):W\subset \mathbb{R}^2 \rightarrow \mathbb{R}^p$ where every $Z_i$ fulfils the same conditions as $Z$.

Considering the increasing domain asymptotic framework and adding also some suitable assumptions, the consistency of his proposal is proved. A bandwidth selection criterion using cross-validation techniques is defined, as well as a dimension reduction tool that allows to handle with high-dimensional covariates.

\cite{Baddeley2012} propose two types of nonparametric intensity estimators, one based on local likelihood and the other on kernel theory. We will focus on the latter, particularly on a kernel intensity estimator for the $\rho$ function with a one-dimensional covariate using weights:

\begin{center}
	\begin{minipage}{0.5\textwidth}
		\begin{equation}\label{Bad_reweighted}
		\hat{\rho}_W(z)=\sum_{i=1}^N\frac{1}{g^\star(Z_i)}K_h(z-Z_i),
		\end{equation}
	\end{minipage}
\end{center}

To obtain the bandwidth parameter $h$, \cite{Baddeley2012} use the common Silverman's rule-of-thumb for density estimation applied to the transformed data.

Our aim in this paper is not only to define a good kernel estimator for the intensity function under assumption \eqref{eq:int_rho}, but to be able to set a theo\-re\-ti\-cal framework in which we could prove its consistency and develop optimal bandwidth methods. In short, we want to be able to characterise the intensity estimator in an proper theoretical framework with all the required tools to be able to apply it in practice. This framework is described in the following section.

\section{A consistent theoretical framework for kernel intensity estimation based on covariates}
In this section we work under the transformed space assuming \eqref{eq:int_rho}, and the point process obtained from the original one, $X$, through the covariate, $Z(X)$, defined in the previous section and detailed in Appendix A.

First of all we need to introduce some definitions and notation. The spatial cumulative distribution function of $Z$ is defined as
\begin{equation*}
G(z)=\frac{1}{|W|}\int_W{1_{\{Z(u)\leq z\}}du},
\end{equation*}
where $|W|$ denotes the area of the region $W \subset \mathbb{R}^2$. Let assume that $G$ has a first derivative $g$, for which we need $Z$ to be differentiable with non-zero gradient and let denote the non-normalised versions by $g^\star(\cdot)=|W|g(\cdot)$ and $G^\star(\cdot)=|W|G(\cdot)$. The results detailed in Appendix A reach to the fact that $\rho g^\star$ is the intensity function of the transformed point process, $Z(X)$. 

Now, following the idea of \cite{CucalaThesis} we use the close relationship between the intensity and the density function. Hence, we can define the following ``artificial'' density function based on the relative density of the transformed point process $Z(X)$:
\begin{equation}\label{eq:dens_int}
f(\cdot)=\frac{\rho(\cdot)g^\star(\cdot)}{m}.
\end{equation}

We propose in this work to take profit of this relationship by firstly estimating the density and then going back to our target problem, that is the intensity estimation, multiplying just by a constant.

Following the pre-established notation, let define the estimator of the re\-la\-ti\-ve density as follows:
\begin{equation}\label{eq:dens_est}
\hat{f}_{h}(z)=g^\star(z) \frac{1}{N}\sum_{i=1}^N\frac{1}{g^\star(Z_i)}K_h\left(z-Z_i\right)1_{\{N\neq 0\}},
\end{equation}
where $K$ is a kernel function and $K_h(\cdot)=\frac{1}{h}K\left(\frac{\cdot}{h}\right)$. This is an estimate of $f$, and once we have it, we can go back to the intensity function just by plug-in and letting $\hat{\lambda}(u)=\hat{\rho}_h(Z(u))$, where $\hat{\rho}_h$ can be replaced by for example \eqref{Bad_reweighted}.

In the following statement we obtain the value of the pointwise mean and variance of $\hat{f}_h$ with the corresponding error rates, as well as its mean squared error (MSE), which is defined as follows:
\[MSE(h,z)=E\left[\left(\hat{f}_h(z)-f(z)\right)^2\right].\]

Hereafter we will establish that our point process $X$ is a nonhomogeneous Poisson point process in $W \subset \mathbb{R}^2$. Although the intensity estimator we propose, as well as the bandwidth selectors, can be applied to non-Poisson processes, the previous assumption is required to prove the consistency of the estimator. 

We also need to introduce some regularity conditions:
\begin{itemize}
	\item[(A.1)]{$\int_{\mathbb{R}}K(z)dz=1$; $\int_{\mathbb{R}}zK(z)dz=0$ and $\mu_2(K):=\int_{\mathbb{R}}z^2K(z)dz<\infty$.}
	\item[(A.2)]{$\lim_{m\to\infty}h=0$ and $\lim_{m\to\infty}\frac{A(m)}{h}=0$, where $A(m):=\mathbb{E}\left[\frac{1}{N}1_{\{N\neq 0\}}\right]$.}
	\item[(A.3)]{$G$ is three times differentiable.} 
	\item[(A.4)]{$z$ is a continuity point of $\rho$.}
	\item[(A.5)]{$\rho$ is three times differentiable.}
\end{itemize}

Notice that we use an infill structure asymptotic framework, which means that the observation region remains fixed while the sample size increases. In this scenario the bandwidth $h$ is considered as a function of the expected sample size, this is,  $h \equiv h(m)$ and hence a sequence of values when $m\to \infty$. 

\begin{theorem}\label{th:mse}
	Under conditions (A.1) to (A.4) we have that:
	\begin{align*}
	E\left[\hat{f}_h(z)\right]&=\frac{g^\star(z)(K_h \circ \rho)(z)}{m}\left(1-e^{-m}\right)\qquad \mbox{and}\\
	Var\left[\hat{f}_h(z)\right]&=A(m)\frac{(g^\star(z))^2}{n}\left(K^2_h \circ \frac{\rho}{g^\star}\right)(z)\\
	&-(A(m)+e^{-2m}-e^{-m})(g^\star(z))^2(K_h\circ \rho)^2(z),
	\end{align*}
	where $\circ$ denotes the convolution between two functions. Moreover, adding condition (A.5) we have:
	\begin{align*}
	MSE(h,z)&=e^{-2m}f^2(z)+(1-e^{-m})^2\frac{h^4}{4}\left(\frac{\rho^{''}(z)g^\star(z)}{m}\right)^2\mu_2^2(K)\nonumber\\
	&-e^{-m}(1-e^{-m})h^2\mu_2(K)\frac{(g^\star(z))^2\rho(z)\rho^{''}(z)}{m^2}+\frac{A(m)}{h}f(z)R(K)\nonumber \\
	&+o(h^2(1-e^{-m})e^{-m})+o(h^4(1-e^{-m})^2)+o\left(\frac{A(m)}{mh}\right),
	\end{align*}
	where $R(K)=\int_{\mathbb{R}}{K^2(z)dz}$.
\end{theorem}

The proofs of this result and the others in the paper are detailed in Appendixes B and C.

Now, defining the mean integrated square error (MISE) as
\begin{equation}\label{MISE}
MISE(h)=E\int{\left(\hat{f}_h(z)-f(z)\right)^2dz},\end{equation}
and denoting by $AMISE$ its asymptotic version, the following result is a consequence of Theorem \ref{th:mse}:

\begin{corollary}\label{th:mise}
	Under conditions (A.1) to (A.3) and (A.5),
	\begin{align*}
	MISE(h)&=e^{-2m}R(f)+(1-e^{-m})^2\frac{h^4}{4}R\left(\frac{\rho^{''}g^\star}{m}\right)\mu_2^2(K)\nonumber \\
	&-e^{-m}(1-e^{-m})h^2\mu_2(K)\int_{\mathbb{R}}\frac{g^\star(z)\rho^{''}(z)f(z)}{m}dz+\frac{A(m)}{h}R(K)\nonumber \\
	&+o(h^2(1-e^{-m})e^{-m})+o(h^4(1-e^{-m})^2)+o\left(\frac{A(m)}{mh}\right) \qquad \mbox{and}
	\end{align*}
	
	\begin{align*}
	AMISE(h)&=(1-e^{-m})^2\frac{h^4}{4}R\left(\frac{\rho^{''}g^\star}{m}\right)\mu_2^2(K)+\frac{A(m)}{h}R(K).
	\end{align*}	
	
	As a consequence, the optimal bandwidth value which minimises $AMISE$ is:
	\begin{equation}\label{eq:hamise}
	h_{AMISE}=\left(\frac{A(m)R(K)}{\mu_2^2(K)(1-e^{-m})^2R\left(\frac{\rho^{''}g^\star}{m}\right)}\right)^{1/5}=\left(\frac{R(K)}{\mu_2^2(K)R(\rho^{''}g^\star)}\frac{A(m)}{(1-e^{-m})^2}\right)^{1/5}.\end{equation}
\end{corollary}

\section{Resampling bootstrap method}
Nonparametric bootstrap procedures have been widely used in different contexts to perform inference and calibrate the distribution of statistics in goodness-of-fit tests. The smooth bootstrap procedure for point processes with covariates proposed in this section is based on the works of \cite{Cao1993} for kernel density estimation, and \cite{Cowling1996}, for the intensity estimation of a Poisson point process.

Let $X_1,\ldots,X_n$ be a realisation of the spatial point process $X$, construct $Z_1,\ldots,Z_n$ the associated realisation of the transformed univariate process, let $\hat{f}_b$ be the density estimator in \eqref{eq:dens_est} and $\hat{\rho}_b$ the estimator defined in the previous section, where $b$ is a pilot bandwidth.

Now, conditional on $Z_1,\ldots,Z_n$, let $N^\ast \sim Poiss\left(\hat{m}\right)$ with $\hat{m}:=\int_{\mathbb{R}}\hat{\rho}_b(z) g^\star(z)dz$, ge\-ne\-ra\-te $n^\ast$ a realisation of this random variable $N^\ast$ and then draw $Z_1^\ast, \ldots, Z_{n^\ast}^\ast$ by sampling randomly with replacement $n^\ast$ times from the distribution with density proportional to $g^\star\hat{\rho}_b$, i.e. $\tilde{f}_b=\frac{\hat{\rho}_b g^\star}{\hat{m}}$.

Denote by $Z^\ast$ the random variable generated by the bootstrap method presented above and from the bootstrap sample define the density estimator as:
\begin{equation}\label{eq:dens_est_boot}
\hat{f}_{h}^\ast(z)=g^\star(z) \frac{1}{N^\ast}\sum_{i=1}^{N^\ast}\frac{1}{g^\star(Z_i^\ast)}K_h\left(z-Z_i^\ast\right)1_{\{N^\ast\neq 0\}},
\end{equation}
hence, using equation \eqref{eq:dens_int} we get the associated estimator of $\rho$:
\[\hat{\rho}_h^\ast(z)=\sum_{i=1}^{N^\ast}\frac{1}{g^\star(Z_i^\ast)}K_h\left(z-Z_i^\ast\right),\]
and then we plug-in it in \eqref{eq:int_rho} to obtain an estimator of $\lambda$.

The following result provides the expression of the mean, variance and mean squared error of $\hat{f}^\ast_h$ under the bootstrap distribution; hereafter we use $E^\ast$, $Var^\ast$ and $MSE^\ast$ to refer to the mean, variance and mean squared error respectively, under the bootstrap distribution.

\begin{theorem}\label{th:mseboot}
	Under conditions (A.1) to (A.4) we get:
	\begin{align*}
	E^\ast\left[\hat{f}_h^\ast(z)\right]&=\frac{g^\star(z)}{\hat{m}}(K_h\circ \hat{\rho}_b)(z)(1-e^{-\hat{m}})\qquad \mbox{and}\\
	Var^\ast\left[\hat{f}^\ast_h(z)\right]&=\frac{(g^\star(z))^2}{\hat{m}}\left(K_h^2 \circ \frac{\hat{\rho}_b}{g^\star}\right)(z)A(\hat{m})-\frac{(g^\star(z))^2}{\hat{m}^2}(K_h\circ \hat{\rho}_b)^2(z)(A(\hat{m})\\
	&+e^{-2\hat{m}}-e^{-\hat{m}}),
	\end{align*}	
	where $A(\hat{m}):=E^\ast\left[\frac{1}{N^\ast}1_{\{N^\ast \neq 0\}}\right]$. Moreover, adding condition (A.5) we have 
	\begin{align}\label{eq:mseboot}
	&MSE^\ast(h,z)=e^{-2\hat{m}}(\tilde{f}_b(z))^2+\frac{h^4}{4}(\hat{\rho}_h^{''}(z))^2\frac{(g^\star(z))^2}{\hat{m}^2}\mu_2^2(K)(1-e^{-\hat{m}})^2\nonumber \\
	&-e^{-\hat{m}}(1-e^{-\hat{m}})h^2\tilde{f}_b(z)\frac{\hat{\rho}_b^{''}(z)g^\star(z)}{\hat{m}}\mu_2(K)+\frac{A(\hat{m})}{h}R(K)+o_P(h^4(1-e^{-\hat{m}})^2)\nonumber \\
	&+o_P(h^2(1-e^{-\hat{m}})e^{-\hat{m}})+o_P\left(\frac{A(\hat{m})}{\hat{m}h}\right);
	\end{align}
	remind that we have defined $\tilde{f}_b(\cdot)=\frac{\hat{\rho}_b (\cdot)g^\star(\cdot)}{\hat{m}}$.
\end{theorem}

In the same way as we have done for Corollary \ref{th:mise}, the integrated and asymptotic version of the $MSE^\ast$ can be easily deduced from the previous result in the next Corollary.
\begin{corollary}\label{th_miseboot}
	Under conditions (A.1) to (A.3) and (A.5),
	\begin{align*}
	&MISE^\ast(h)=e^{-2\hat{m}}R(\tilde{f}_b)+\frac{h^4}{4}R\left(\frac{\hat{\rho}_b^{''}g^\star}{\hat{m}}\right)\mu_2^2(K)(1-e^{-\hat{m}})^2\\
	&-e^{-\hat{m}}(1-e^{-\hat{m}})h^2\mu_2(K)\int{\frac{\tilde{f}_b(z)\hat{\rho}_b^{''}(z)g^\star(z)}{\hat{m}}dz}+\frac{A(\hat{m})}{h}R(K) \\
	&+o(h^4(1-e^{-\hat{m}})^2)+o_P(h^2(1-e^{-\hat{m}})e^{-\hat{m}})+o_P\left(\frac{A(\hat{m})}{\hat{m}h}\right)  \qquad \mbox{and}
	\end{align*}
	\begin{align*}
	&AMISE^\ast(h)=\frac{h^4}{4}R\left(\frac{\hat{\rho}_b^{''}g^\star}{\hat{m}}\right)\mu_2^2(K)(1-e^{-\hat{m}})^2+\frac{A(\hat{m})}{h}R(K).
	\end{align*}
	Therefore the asymptotic expression of the optimal bootstrap bandwidth is:
	
	\begin{equation}\label{eq:eq_hamiseboot}
	h_{AMISE^\ast}=\left(\frac{A(\hat{m})R(K)}{\mu_2^2(K)(1-e^{-\hat{m}})^2R\left(\frac{\hat{\rho}_b^{''}g^\star}{\hat{m}}\right)}\right)^{1/5},
	\end{equation}
	which is a plug-in version of \eqref{eq:hamise}.
\end{corollary}

All these results above lead to the following Corollary.
\begin{corollary}Under assumptions (A.1) to (A.4) $MISE^\ast$ and $AMISE^\ast$ are consistent estimators of $MISE$ and $AMISE$, respectively.
\end{corollary}

\begin{remark}
	The theory developed in this section is restricted to the context of spatial point processes, i.e., processes in $\mathbb{R}^2$. A generalisation to the $\mathbb{R}^p$ space can be done without much  more complexity but the increased dimension of the involved parameters, however it does not seem to be of great practical value. 
\end{remark}

\section{Data-driven bandwidth selection}
In this section we describe two new bandwidth selection methods for the intensity estimation based on \eqref{eq:dens_est}. These methods consist of adaptations of common selectors in the field of density estimation that have not yet been defined nor implemented in the point process framework. We propose a Normal scale rule (rule-of-thumb) and a bootstrap selector derived from the consistent resampling bootstrap procedure detailed in the previous section. All these proposals are based on estimating the infeasible optimal expression \eqref{eq:hamise} in which the unknown elements are $m$, $A(m)$ and $\rho^{''}$.

\subsection{Rule-of-thumb for bandwidth selection}
The basis idea of this method is the same as in \cite{Silverman1986}: we assume that the underlying density \eqref{eq:dens_int} is Normal, $N(\mu,\sigma)$, with the parameters being estimated from the data, and in this way we replace the unknown values in \eqref{eq:hamise}.

In the point processes framework the computation is slightly different from the one used in the context of density estimation, because here the density is only a feature to get the intensity. To begin with, we have to remark that in our context, \eqref{eq:hamise} has some other unknown elements apart from $f$, such as $m$ and $A(m)$. The first one is the expected number of points, that in practice can be estimated by the sample size $n$, and the second one by $1/n$.

The only unknown element left is $\rho^{''}$. However, assuming that $f=\frac{\rho g^\star}{m}$ is Normal, we can derive that
\[\rho{''}(z)=m\left(\frac{f^{''}(z)}{g^\star(z)}-\frac{2f^{'}(z)(g^\star(z))^{'}}{(g^\star(z))^2}-\frac{f(z)(g^\star(z))^{''}}{(g^\star(z))^2}+\frac{2f(z)((g^\star(z))^{'})^2}{(g^\star(z))^3}\right),\]
and then compute $R\left(\frac{\rho^{''}g^\star}{m}\right)$ using numerical integration methods. Replacing all those estimates in \eqref{eq:hamise} we have the rule-of-thumb bandwidth selector that we will denote by $\hat{h}_{\textup{RT}}$.

\subsection{Bootstrap for bandwidth selection}
The asymptotic expression of the optimal bootstrap bandwidth can be considered to
derive a consistent bandwidth estimate. \cite{Cao1993} suggested such approach for kernel
density estimation with complete data and \cite{BorrajoLengthBiased} detail the result for length-biased data, as well as some remarks to extend it to general weighted data. 

In the expression \eqref{eq:eq_hamiseboot} that we use to build this selector,some quantities need to be computed: $\hat{m}$, $A(\hat{m})$ and $R(\frac{\hat{\rho}_b^{''}g^\star}{\hat{m}})$. The first two can be easily calculated through numerical integration methods such as Simpson's rule, while the last one requires some more effort.

The main challenge in the estimation of $R(\frac{\hat{\rho}_b^{''}g^\star}{\hat{m}})$ is to obtain an appropriate value for the pilot bandwidth $b$. Regarding \cite{Cao1993} and \cite{BorrajoLengthBiased} we can assume that the order of that bandwidth in our context is $m^{-1/7}$, and that the constant has a slight influence on the final result. Hence we propose to use as pilot bandwidth a re-scaled version of the rule-of-thumb previously defined:

\begin{equation*}
\hat{b}=\frac{m^{-1/5}}{m^{-1/7}}\hat{h}_{RT}.
\end{equation*}
Obviously in practice we do not know the value of $m$, so we use the best approximation we can have which is the sample size of the corresponding realization of the point process.

Then, the bootstrap bandwidth we propose is:
\begin{equation*}
\hat{h}_{\mbox{Boot}}=\left(\frac{A(\hat{m})R(K)}{\mu_2^2(K)(1-e^{-\hat{m}})^2R(\frac{\hat{\rho}_{\hat{b}}^{''}g^\star}{\hat{m}})}\right)^{1/5}.
\end{equation*}

\section{Finite sample study}

In this section we perform a simulation study to analyse the behaviour of the methods proposed in this paper. Firstly we analyse the performance of the intensity estimator defined in  (\ref{Bad_reweighted}),  using our two bandwidth selection pro\-po\-sals: the rule-of-thumb, $\hat{h}_{\textup{RT}}$, and the bootstrap method, $\hat{h}_{\textup{Boot}}$. We compare these with the only bandwidth selector that has been previously proposed by \cite{Baddeley2012} for this intensity estimator, which is the common Silverman's rule-of-thumb for density estimation denoted here as $\hat{h}_{\textup{Silv}}$. We also added in this comparison a least-squares cross-validation bandwidth selection method defined for our estimator in the same way as it is done for common density estimation, i.e., using least-squares cross-validation and this will be denoted by $\hat{h}_{\textup{CV}}$. Secondly we extend the comparison to the intensity estimator proposed by \cite{Guan2008} for which also a cross-validation bandwidth selector was suggested. 

%

\begin{figure}[H]
	\centering
	\includegraphics[scale=0.32,angle=90]{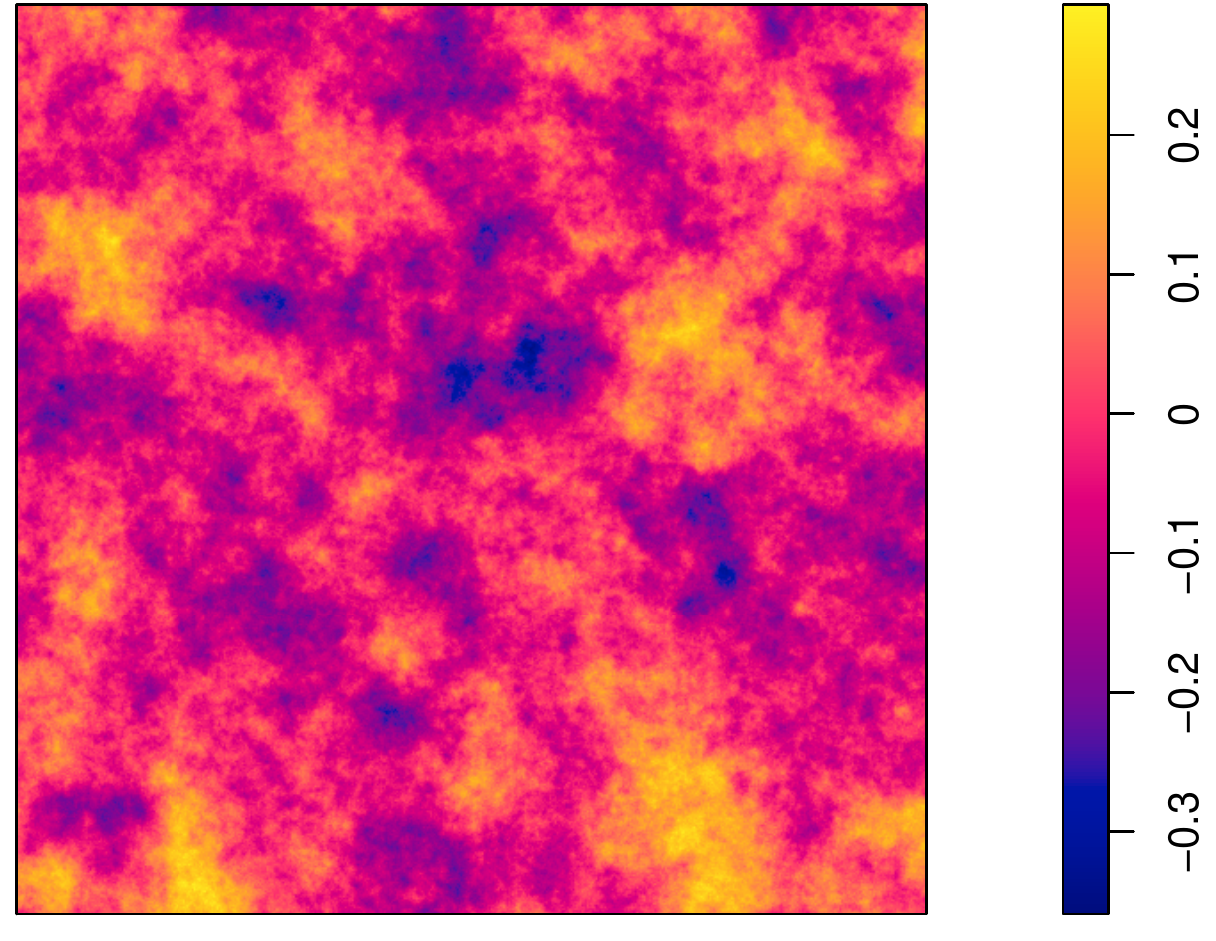}
	\hspace*{1cm}
	\includegraphics[scale=0.32,angle=90]{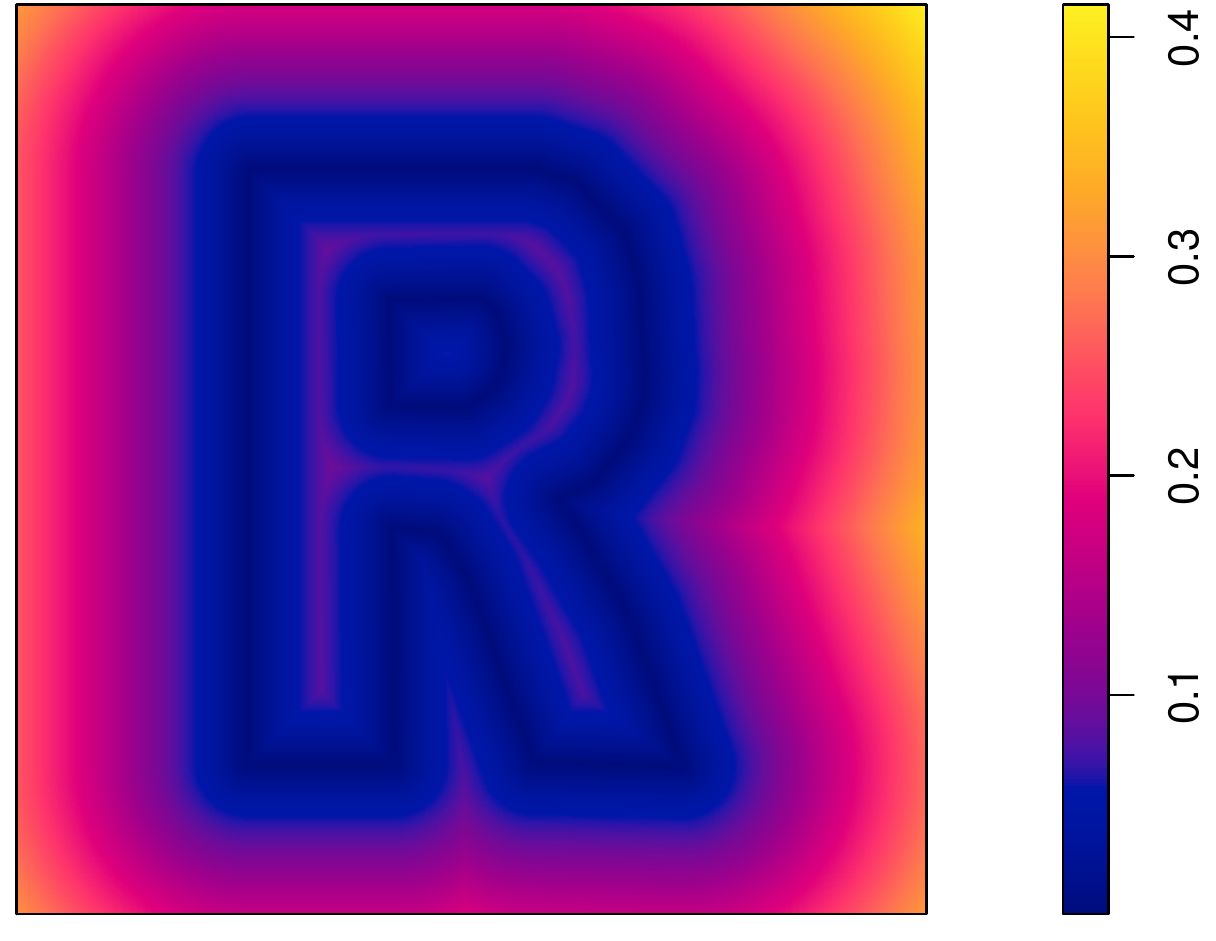}
	\caption{Plot of the covariates used in the three models, $Z1$ on the left and dR on the right.}	\label{fig:covariates}
\end{figure}

We have chosen three different models, all under assumption \eqref{eq:int_rho}. These models are nonhomogeneous Poisson point processes in the unit square with intensity function given by 
\begin{equation*}\lambda(u)=exp(\beta_0+\beta_1Z(u)), \quad u \in W=[0,1]\times[0,1],
\end{equation*}
where $\beta_0$ and $\beta_1$ are known parameters, and $Z$ denotes a covariate.
The three models are constructed using two different  covariates that are shown in Fig. \ref{fig:covariates}. The first covariate on the left, $Z1$, is a realisation of a Gaussian random field, with zero mean and exponential covariance structure with parameters $\sigma=0.1$ and $s=0.1$, so the covariance function is given by $C(r)=\sigma^2exp(-r/s)$. The second covariate on the right, dR, is a rescaled version into the unit square of the ``distance to letter R'' defined in \cite{Baddeley2012}. 

\vspace*{-0.5cm}

The three models are defined by the following three intensity functions:
\begin{eqnarray*}
	\lambda_1(u)&=&\exp{(6+4Z1(u))} \\
	\lambda_2(u)&=&\exp{(6+4(Z1+e1)(u))}\\
	\lambda_{3}(u)&=&\exp{(5-3\mbox{dR}(u))} 
\end{eqnarray*}\vspace*{-0.5cm}

\begin{figure}[H]
	\centering
	\includegraphics[scale=0.32,angle=90]{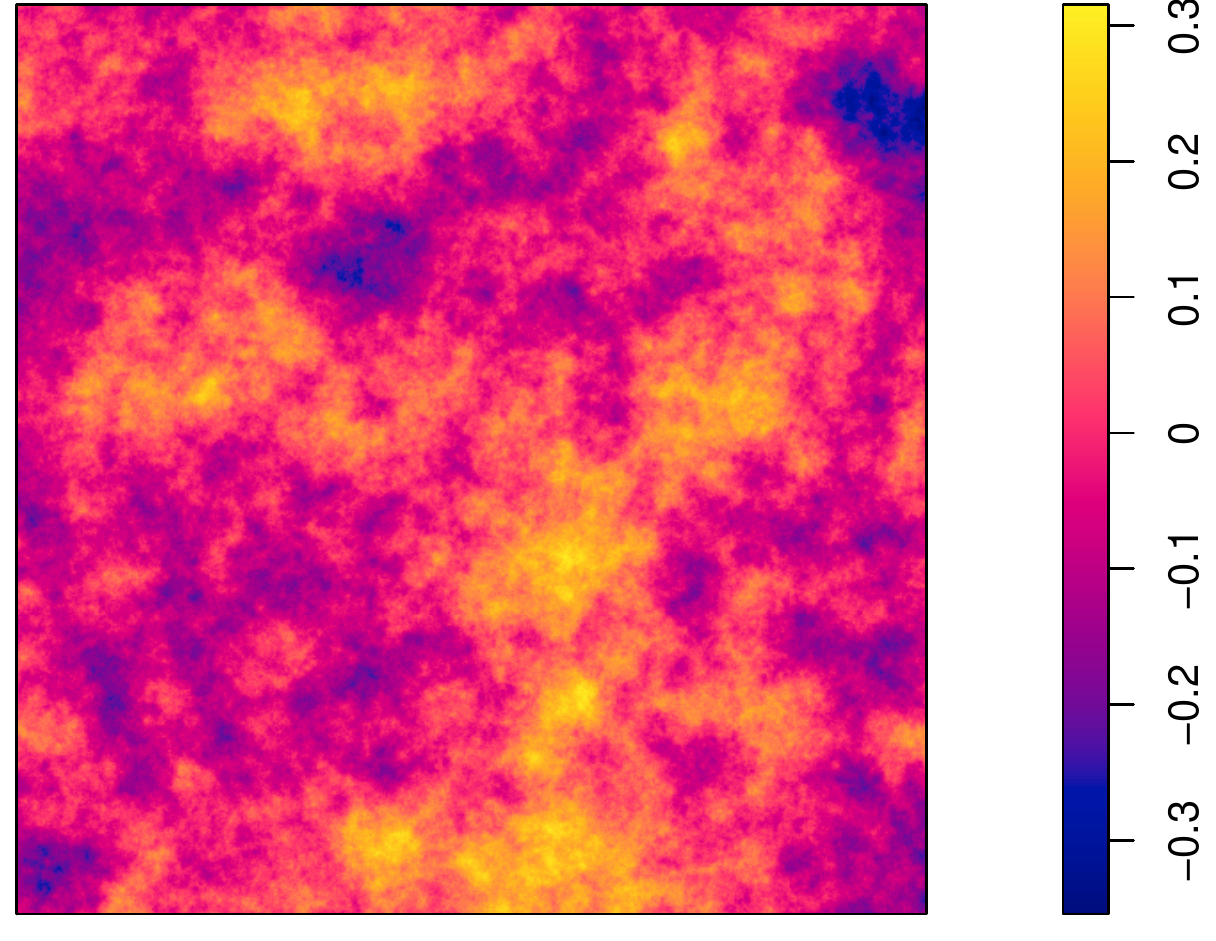}
	\caption{Plot of the error term used to perturb the information given to the estimator by the real covariate.}	\label{fig:e1}
\end{figure}

\begin{figure}[H]
	\begin{center}
		\subfigure[Model 1]{
			\includegraphics[scale=0.26,angle=90]{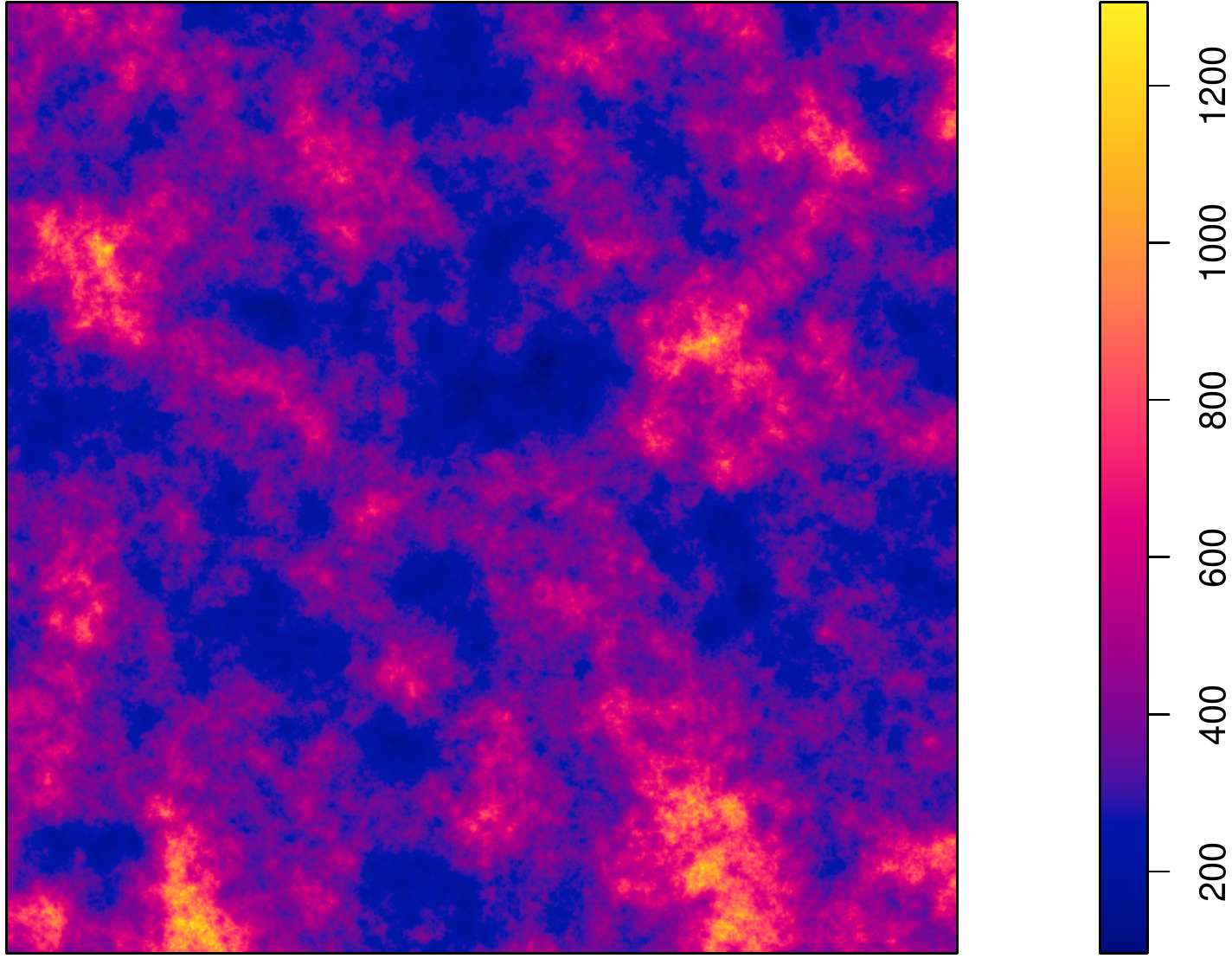}}\hspace*{0.1cm}
		\subfigure[Model 2]{
			\includegraphics[scale=0.26,angle=90]{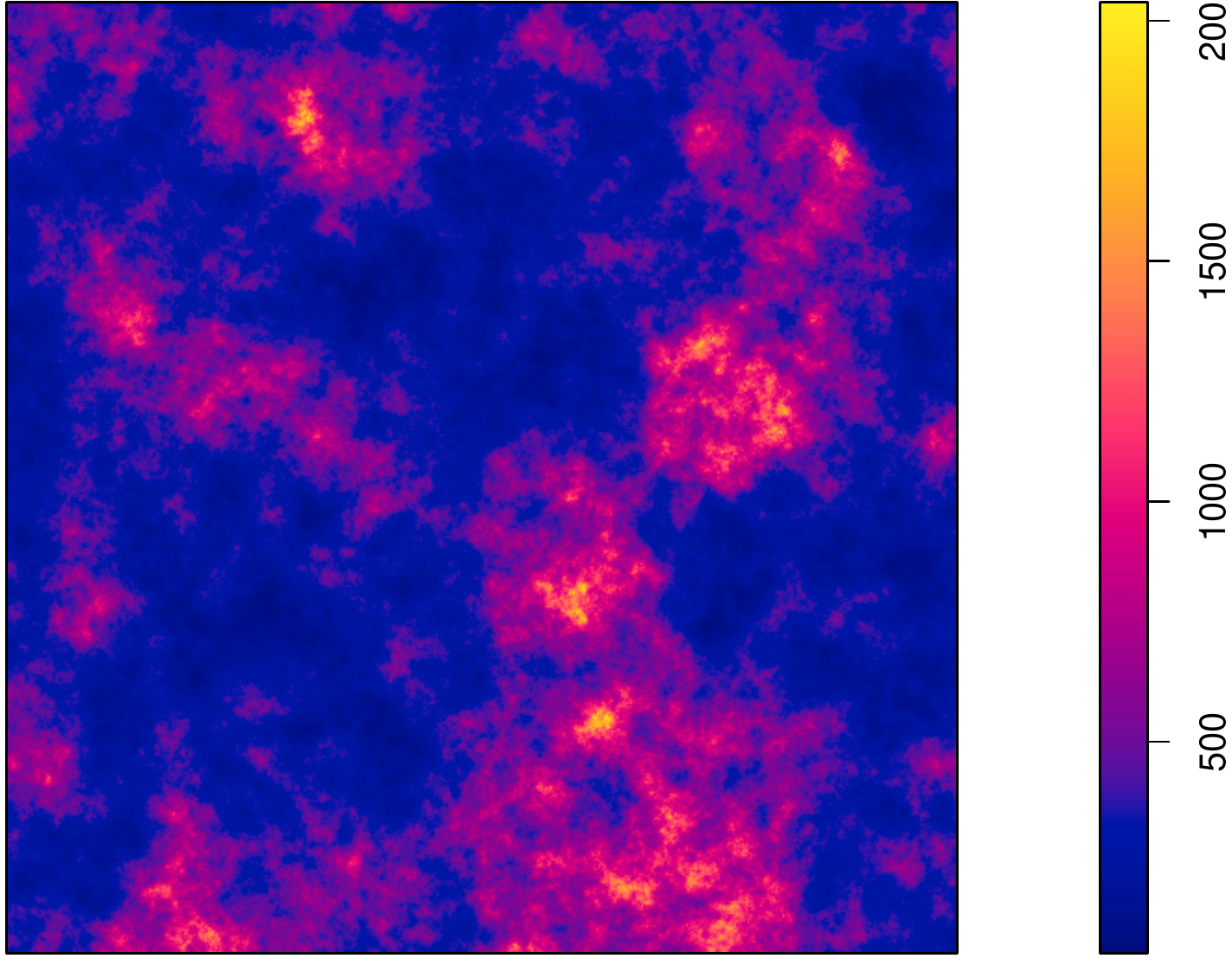}}\hspace*{0.1cm}		
		\subfigure[Model 3]{
			\includegraphics[scale=0.26,angle=90]{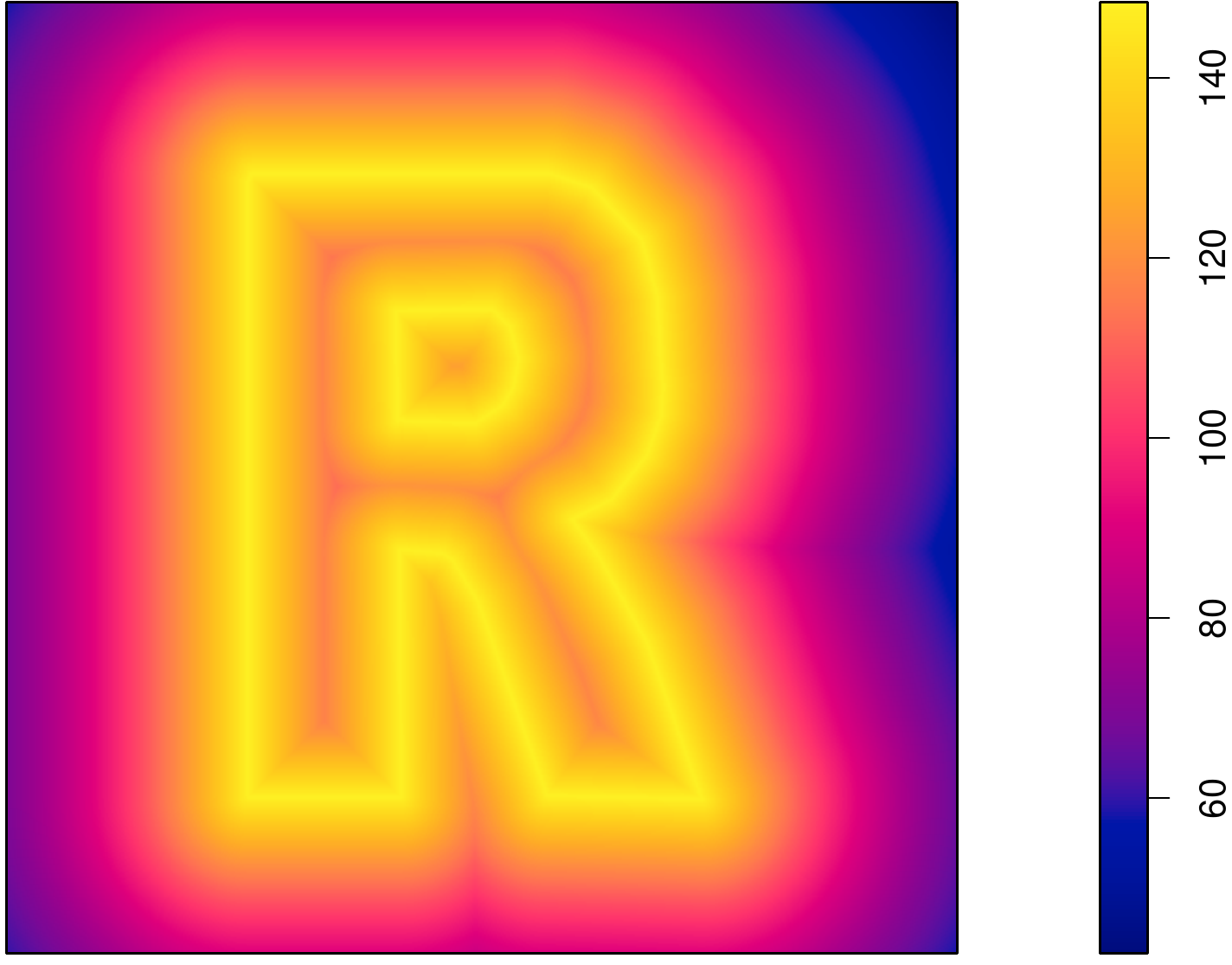}}
		\caption{Intensity functions characterising the three simulated models.}	\label{fig:models}
	\end{center}
\end{figure}

The second model includes an error term to perturb the covariate, in order to see how the estimators perform using only partial information on the real covariate generating the process. We have generated this error term considering a different realisation of the same Gaussian random field we have defined for $Z1$. A plot of this error term is shown in Fig. \ref{fig:e1}, and the three intensity functions are represented in Fig. \ref{fig:models}.\vspace*{-0.3cm}

\vspace*{-1cm}

We have simulated 500 realisations for each model and different expected sample sizes covering a wide range of possibilities, $m=50, 100, 250 \mbox{ and } 500$. Notice that the underlying model generating these samples is not exactly the same,  we need to rescale the intensity function to guarantee that the mean number of events in the unit square is $m$. 
From the simulated samples, we have evaluated the performance of the intensity estimator in (\ref{Bad_reweighted}) with different  bandwidth choices through three error criteria, $e_1$, $e_2$ and $e_3$, defined below.

Our final aim is to estimate the intensity function and so we want to show the performance of the resulting intensity estimators for each bandwidth selector. For two-dimensional intensity estimates we consider the relative integrated squared error defined as:
\begin{equation} \label{ISErel}
\mbox{ISE}_{\mbox{rel}}=\int_W{\left(\frac{\hat{\lambda}(u)-\lambda(u)}{\lambda(u)}\right)^2du},
\end{equation}
and define the first two performance measures as:
\begin{equation*}
\begin{array}{c}
e_1=\textup{mean}\left(\textup{ISE}_{\mbox{rel}}(\hat{h})\right) \ \ \ \mbox{and} \ \ \ e_2=\textup{std}\left(\textup{ISE}_{\mbox{rel}}(\hat{h})\right) ,
\end{array}
\end{equation*}
which represent the average relative error and the variability around, respectively. 
On the other hand, our bandwidth selectors aim to estimate the infeasible optimal bandwidth that minimises the $MISE$ criterion defined in (\ref{MISE}). So it is natural to consider such infeasible value as a benchmark in our simulations, and measure how close our estimates are from such value. This motivates our third performance measure that is the relative bias of the bandwidth selectors defined as:
\begin{equation*}
\begin{array}{c}
e_3=\textup{mean}\left((\hat{h}-\hat{h}_{\textup{MISE}})/\hat{h}_{\textup{MISE}}\right),
\end{array}
\end{equation*}
where $\hat{h}_{\textup{MISE}}$ is the minimiser of the Monte Carlo approximation (based on the 500 simulated samples) of criterion \eqref{MISE}.  
The simulation results are summarised in Tables \ref{tb1}, \ref{tb2} and \ref{tb3}.

\renewcommand{\arraystretch}{1.2}
\begin{table}[H]
	\caption{Simulation results for Model 1. Performance measures  $e_1$ to $e_3$ are reported for the intensity estimator with bandwidths $\hat{h}_{\textup{MISE}}$, $\hat{h}_{\textup{Silv}}$, $\hat{h}_{\textup{RT}}$, $\hat{h}_{\textup{Boot}}$ and $\hat{h}_{\textup{CV}}$.} \label{tb1}
	\centering
	\begin{tabular}{lccccc} \hline
		& \multicolumn{5}{c}{\textit{Model 1}}  \\
		& $\hat{h}_{\textup{MISE}}$ & $\hat{h}_{\textup{Silv}}$ & $\hat{h}_{\textup{RT}}$ &  $\hat{h}_{\textup{Boot}}$ & $\hat{h}_{\textup{CV}}$  \\ \hline
		& \multicolumn{5}{c}{$m=50$} \\
		$e_1$ & 0.0662 &  0.1222 & 0.0881 & 0.0656 & 0.2688 \\ 
		$e_2$ & 0.0735 &  0.1089 & 0.0848 & 0.0701 & 0.0699 \\
		$e_3$ &    --- & -0.6132 &-0.4352 & 0.0533 & 6.2584 \\	
		& \multicolumn{5}{c}{$m=100$} \\
		$e_1$ & 0.0388 &  0.0706 & 0.0518 & 0.0391 & 0.2665 \\
		$e_2$ & 0.0345 &  0.0494 & 0.0391 & 0.0342 & 0.0499 \\
		$e_3$ &    --- & -0.6238 &-0.4577 & 0.0686 & 6.5941\\
		& \multicolumn{5}{c}{$m=200$} \\
		$e_1$ & 0.0248 &  0.0443 & 0.0332 & 0.0250 & 0.2731 \\
		$e_2$ & 0.0220 &  0.0339 & 0.0268 & 0.0225 & 0.0347 \\
		$e_3$ &    --- & -0.6274 &-0.4706 & 0.0859 & 8.1597\\
		& \multicolumn{5}{c}{$m=500$} \\
		$e_1$ & 0.0127 &  0.0231 & 0.0173 & 0.0131 & 0.2768\\
		$e_2$ & 0.0091 &  0.0159 & 0.0120 & 0.0093 & 0.0201\\
		$e_3$ &    --- & -0.6254 &-0.4690 & 0.1566 & 10.1974 \\ \hline
	\end{tabular}
\end{table}	

\renewcommand{\arraystretch}{1.2}
\begin{table}[H]
	\caption{Simulation results for Model 2. Performance measures  $e_1$ to $e_3$ are reported for the intensity estimator with bandwidths $\hat{h}_{\textup{MISE}}$, $\hat{h}_{\textup{Silv}}$, $\hat{h}_{\textup{RT}}$, $\hat{h}_{\textup{Boot}}$ and $\hat{h}_{\textup{CV}}$.}\label{tb2}
	\centering
	\begin{tabular}{lccccc} \hline
		& \multicolumn{5}{c}{\textit{Model 2}}  \\
		& $\hat{h}_{\textup{MISE}}$ & $\hat{h}_{\textup{Silv}}$ & $\hat{h}_{\textup{RT}}$ &  $\hat{h}_{\textup{Boot}}$ & $\hat{h}_{\textup{CV}}$ \\ \hline
		& \multicolumn{5}{c}{$m=50$} \\
		$e_1$ & 0.3070 &  0.3414 & 0.3115 & 0.3045 & 0.2708\\ 
		$e_2$ & 0.1451 &  0.1567 & 0.1436 & 0.1470 & 0.0501\\
		$e_3$ &    --- & -0.4386 &-0.2109 & 0.4183 & 0.2839\\	
		& \multicolumn{5}{c}{$m=100$} \\
		$e_1$ & 0.2574 &  0.2830 & 0.2708 & 0.2810 & 0.2660\\
		$e_2$ & 0.1270 &  0.1109 & 0.1061 & 0.1092 & 0.0329\\
		$e_3$ &    --- & -0.9252 &-0.8985 &-0.8111 & 0.4187\\
		& \multicolumn{5}{c}{$m=200$} \\
		$e_1$ & 0.2269 &  0.2514 & 0.2453 & 0.2617 & 0.2685\\
		$e_2$ & 0.0717 &  0.0727 & 0.0660 & 0.0665 & 0.0247\\
		$e_3$ &    --- & -0.9334 &-0.9099 &-0.8290 & 0.5097\\
		& \multicolumn{5}{c}{$m=500$} \\
		$e_1$ & 0.2065 &  0.2281 & 0.2264 & 0.2421 & 0.2710 \\
		$e_2$ & 0.0313 &  0.0363 & 0.0355 & 0.0368 & 0.0157 \\
		$e_3$ &    --- & -0.9419 &-0.9252 &-0.8540 & 0.5688\\ \hline
	\end{tabular}
\end{table}

\renewcommand{\arraystretch}{1.2}
\begin{table}[H]
	\caption{Simulation results for Model 3. Performance measures  $e_1$ to $e_3$ are reported for the intensity estimator with bandwidths $\hat{h}_{\textup{MISE}}$, $\hat{h}_{\textup{Silv}}$, $\hat{h}_{\textup{RT}}$, $\hat{h}_{\textup{Boot}}$ and $\hat{h}_{\textup{CV}}$. }\label{tb3}
	\centering
	\begin{tabular}{lccccc} \hline
		& \multicolumn{5}{c}{\textit{Model 3}}  \\
		& $\hat{h}_{\textup{MISE}}$ & $\hat{h}_{\textup{Silv}}$ & $\hat{h}_{\textup{RT}}$ & $\hat{h}_{\textup{Boot}}$ & $\hat{h}_{\textup{CV}}$\\ \hline
		& \multicolumn{5}{c}{$m=50$} \\
		$e_1$ & 0.0970 &  0.1173 & 0.1012 & 0.1001 & 0.1798\\ 
		$e_2$ & 0.0845 &  0.1013 & 0.0877 & 0.0855 & 0.0621\\
		$e_3$ &    --- & -0.3810 &-0.1910 & 0.2486 & 2.3346\\	
		& \multicolumn{5}{c}{$m=100$} \\
		$e_1$ & 0.0669 &  0.0735 & 0.0679 & 0.0704 & 0.1689\\
		$e_2$ & 0.0419 &  0.0507 & 0.0441 & 0.0396 & 0.0407\\
		$e_3$ &    --- & -0.2854 &-0.1001 & 0.3110 & 3.4686\\
		& \multicolumn{5}{c}{$m=200$} \\
		$e_1$ & 0.0446 & 0.0460 & 0.0447 & 0.0487 & 0.1632\\
		$e_2$ & 0.0237 & 0.0251 & 0.0234 & 0.0227 & 0.0293\\
		$e_3$ &    --- &-0.1730 & 0.0243 & 0.4022 & 4.8131\\
		& \multicolumn{5}{c}{$m=500$} \\
		$e_1$ & 0.0244 &  0.0244 & 0.0251 & 0.0283 & 0.1587\\
		$e_2$ & 0.0094 &  0.0093 & 0.0088 & 0.0090 & 0.0183\\
		$e_3$ &    --- & -0.0003 & 0.2245 & 0.5455 & 0.5229\\ \hline
	\end{tabular}
\end{table}

An overview of the values in Tables \ref{tb1}, \ref{tb2} and \ref{tb3} indicates that in general, the bootstrap bandwidth seems to perform better than the others in most of the cases, and when this does not occur, our procedure is still competitive. Any of the other rule-of-thumb are not far away from it, even though the one specifically designed for spatial point processes show a slightly better behaviour than the Silverman's, specially for small sample sizes. The cross-validation criteria is only competitive in Model 2, where the covariate information provided to the estimator is contamined with an error term.

In terms of variability, measured by criteria $e_2$, the four compared methods are similar, even though the bootstrap estimates shows in general smaller values.

Looking at the bias of bandwidth estimates, measured through criterion $e_3$, for Model 1, the bootstrap bandwidth selector outperforms far more better than the others; for Model 2, the cross-validation reaches the smaller values, while for Model 3 the rule-of-thumb outperforms the others. Notice also that the rule-of-thumb and Silverman's procedures show the bias in the same direction, to be more specific all of them tend to choose smaller bandwidths than the optimal one. And also remark that this measure is not completely fair for the cross-validation criterion because its objective is not the optimal bandwidth in terms of MISE but in terms of ISE, which might also explain the high values shown in this measure.

To complete our analysis, we have carried out a parallel simulation study to compare our proposals to the competitor described in \cite{Guan2008}. 
The simulation results are summarised in Table \ref{tb4}, where we report the performance measures $e_1$ and $e_2$, for the intensity estimator proposed by Guan and our proposal in this paper. For Guan's intensity estimator we have considered two bandwidth choices: the (Monte Carlo approximated) optimal MISE bandwidth, $\hat{h}_{\textup{MISE}_\textup{Guan}}$, which is considered as a benchmark for this estimator,  and the practical least-squares cross-validation bandwidth, $\hat{h}_{\textup{CV}_\textup{Guan}}$, proposed in his paper. For our proposal we have considered the intensity estimator with the benchmark $\hat{h}_{\textup{MISE}}$ defined above, and our bootstrap bandwidth selector, $\hat{h}_{\textup{Boot}}$. 

\begin{table}[H]
	\caption{Comparison with Guan's estimator. Measures $e_1$ and $e_2$ are reported for our intensity estimator with bandwidths $\hat{h}_{\textup{MISE}}$ and $\hat{h}_{\textup{Boot}}$, and compared to Guan's estimator with optimal-MISE bandwidth, $\hat{h}_{\textup{MISE}_{\textup{Guan}}}$, and a cross-validation estimate, $\hat{h}_{\textup{CV}}$.}\label{tb4}
	\centering
	\resizebox*{\textwidth}{!}{\begin{tabular}{lcccccccc} \hline
			& \multicolumn{4}{c}{$m=50$} & \multicolumn{4}{c}{$m=100$}  \\ \hline
			& $\hat{h}_{\textup{MISE}_{\textup{Guan}}}$ & $\hat{h}_{\textup{CV}_\textup{Guan}}$ & $\hat{h}_{\textup{MISE}}$ & $\hat{h}_{\textup{Boot}}$ & $\hat{h}_{\textup{MISE}_{\textup{Guan}}}$  & $\hat{h}_{\textup{CV}_\textup{Guan}}$ & $\hat{h}_{\textup{MISE}}$ & $\hat{h}_{\textup{Boot}}$  \\ \hline 
			\multicolumn{9}{c}{ \textit{Model 1}} \\
			$e_1$ & 0.0815 & 0.1855 & 0.0662 &  0.0656 & 0.0481 & 0.1885 & 0.0388 & 0.0391 \\ 
			$e_2$ & 0.0786 & 0.1367 & 0.0735 &  0.0701 & 0.0412 & 0.1034 & 0.0345 & 0.0342 \\
			\multicolumn{9}{c}{\textit{Model 2}} \\
			$e_1$ & 0.3316 & 0.4479 & 0.3070 &  0.3045 & 0.2777 & 0.4027 & 0.2574 &  0.2810 \\ 
			$e_2$ & 0.1804 & 0.2474 & 0.1451 &  0.1470 & 0.1092 & 0.1756 & 0.1270 &  0.1092 \\
			\multicolumn{9}{c}{\textit{Model 3}} \\
			$e_1$ & 0.0518 & 0.0737 & 0.0970 & 0.1001 & 0.0309 & 0.0442 & 0.0669 & 0.0704 \\ 
			$e_2$ & 0.0530 & 0.0822 & 0.0845 & 0.0855 & 0.0305 & 0.0457 & 0.0419 & 0.0396 \\
			&&&&&&&&\\\hline 
			& \multicolumn{4}{c}{$m=200$} & \multicolumn{4}{c}{$m=500$}  \\ \hline
			& $\hat{h}_{\textup{MISE}_{\textup{Guan}}}$ & $\hat{h}_{\textup{CV}_\textup{Guan}}$ & $\hat{h}_{\textup{MISE}}$ & $\hat{h}_{\textup{Boot}}$ & $\hat{h}_{\textup{MISE}_{\textup{Guan}}}$  & $\hat{h}_{\textup{CV}_\textup{Guan}}$ & $\hat{h}_{\textup{MISE}}$ & $\hat{h}_{\textup{Boot}}$  \\ \hline 
			\multicolumn{9}{c}{ \textit{Model 1}} \\
			$e_1$ & 0.0288 & 0.2358 & 0.0248 & 0.0250 & 0.0147 & 0.2407 & 0.0127 & 0.0131 \\
			$e_2$ & 0.0243 & 0.0892 & 0.0220 & 0.0225 & 0.0097 & 0.0640 & 0.0091 & 0.0093 \\
			\multicolumn{9}{c}{\textit{Model 2}} \\
			$e_1$ & 0.2482 & 0.3804 & 0.2269 & 0.2617 & 0.2283 & 0.3593 & 0.2065 & 0.2421 \\
			$e_2$ & 0.0707 & 0.1303 & 0.0717 & 0.0665 & 0.0393 & 0.0825 & 0.0313 & 0.0368 \\
			\multicolumn{9}{c}{\textit{Model 3}} \\
			$e_1$ & 0.0181 & 0.0356 & 0.0446 & 0.0487 & 0.0092 & 0.0227 & 0.0244 & 0.0238 \\
			$e_2$ & 0.0152 & 0.0274 & 0.0237 & 0.0227 & 0.0064 & 0.0155 & 0.0094 & 0.0090 \\
	\end{tabular}}
\end{table}		
The results in Table \ref{tb4} are derived from the same 500 simulated samples considered in previous tables for the four sample sizes $m=50, 100, 200$ and $500$. We want to point out the computational burden of performing cross validation for Guan's estimator, along with the  numerical integration that this estimator requires. It seems to be a drawback of this approach that is well known for cross-validation methods. Another issue that we have observed in this context is the non existence of a global minimum for the cross-validation score in a number of samples. In our simulations this issue has leaded to the selection of the bandwidth at the boundary of the minimisation interval, which has occurred between 45 and 213 times out of 500, depending on the model and the sample sizes (it is more likely for smaller sizes).
If we ignore these issues of the practical proposal by Guan, we can see from the numbers reported in Table \ref{tb4} that our proposal performs considerably better than Guan's for the two first simulated models. In the best situation for the intensity estimators, i.e., calculating the intensity estimators with the infeasible benchmarks, $\hat{h}_{\textup{MISE}}$ and $\hat{h}_{\textup{MISE}_\textup{Guan}}$, our estimator achieves smaller relative errors with slightly lower  variability. Considering the practical bandwidth choices for each estimator our bootstrap approach clearly beats the cross-validation method. On the other hand, Model 3 seems to be a good scenario for Guan's estimator and his practical cross-validation method, but even in this case our bootstrap proposal is still competitive. 

%
%
%
%
%
%
%
%
%
%

\section{Canadian wildfires}
Forest fires are one of the most important natural disturbances since the last Ice Age and they represent a huge social and economic problem. Canada has quite a long tradition on recording information about their wildfires; and also studies from many different perspectives have been carried out: \cite{CanadaFires_Clouds}, \cite{CanadaFires_HighLatitudeCooling}, \cite{CanadaFires_SmokeMediterraneanSea}, \cite{CanadaFires_Meteorological}. It is quite well known that fire activity in Canada mostly relies on meteorological elements such as long periods without rain and high temperatures.

\begin{figure}[H]
	\centering
	\subfigure{\resizebox*{.4\textwidth}{!}{\includegraphics{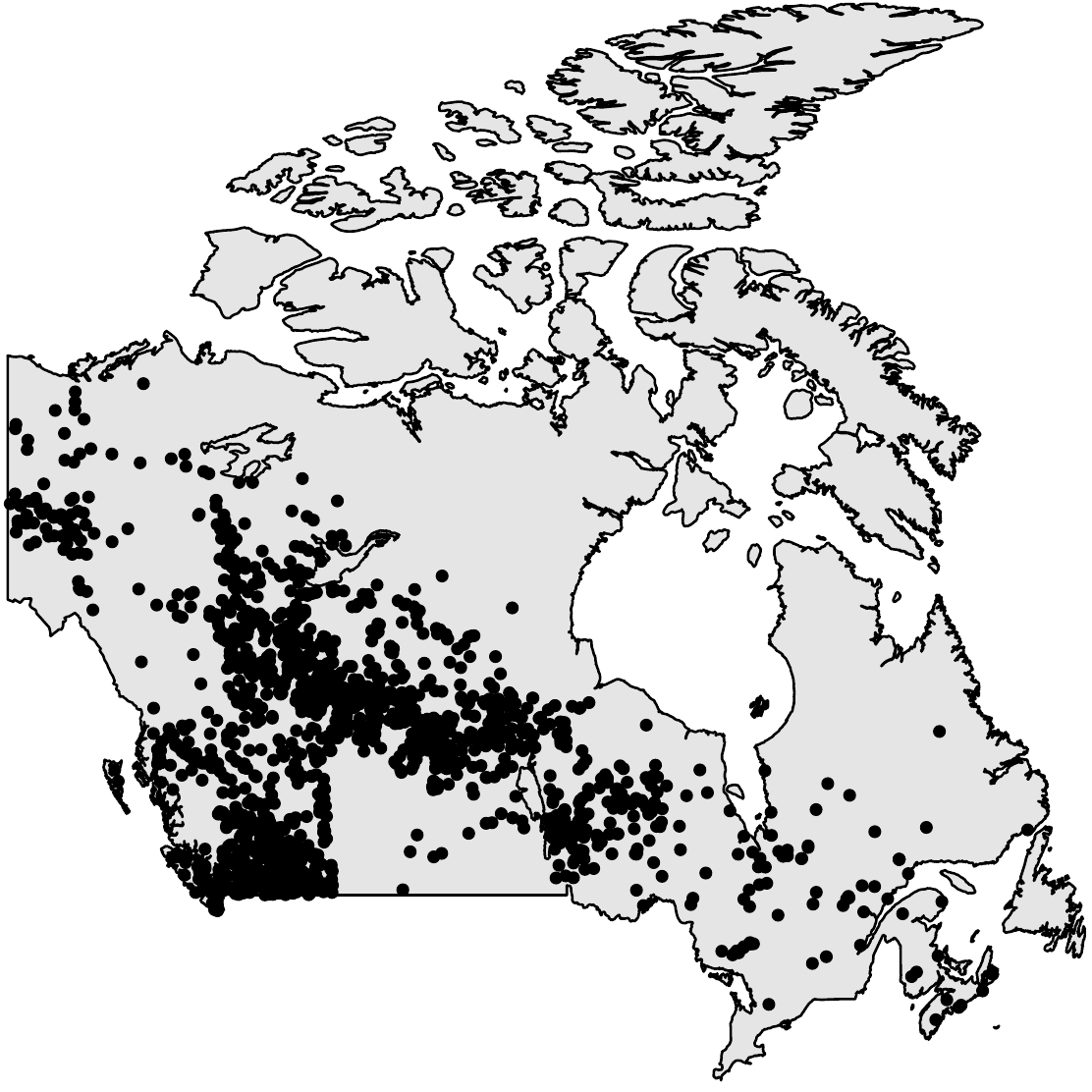}}}
	\hspace*{0.5cm}		\subfigure{\resizebox*{.4\textwidth}{!}{\includegraphics{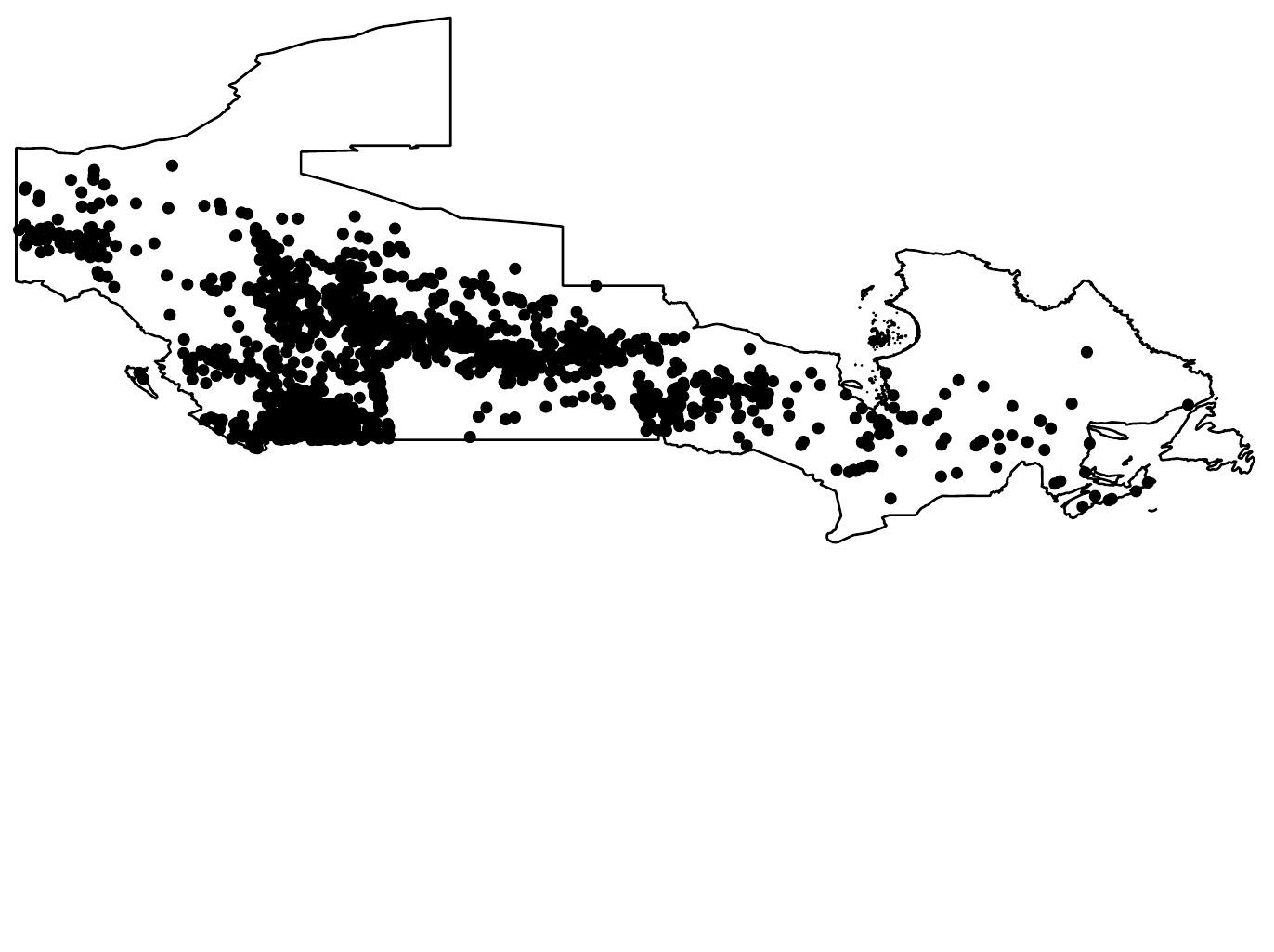}}}
	\caption{Wildfires in Canada during June 2015, over the whole country (left) and only on the observation region (right).}\label{fg:incjun2015}
\end{figure}

It is important to note that for inferential purposes we have removed two regions (Northwest Territories and Nunavut) from the whole observation region (Canada) because there are no fires registered on those regions and we cannot do any inference with such a lack of information.

\begin{figure}[H]
	\centering
	\subfigure[Third quartile of the temperature registered in June 2015 after a Gaussian smoothing with $\sigma=2$ (in Celsius degrees)]{
		\resizebox*{.45\textwidth}{!}{\includegraphics[angle=90]{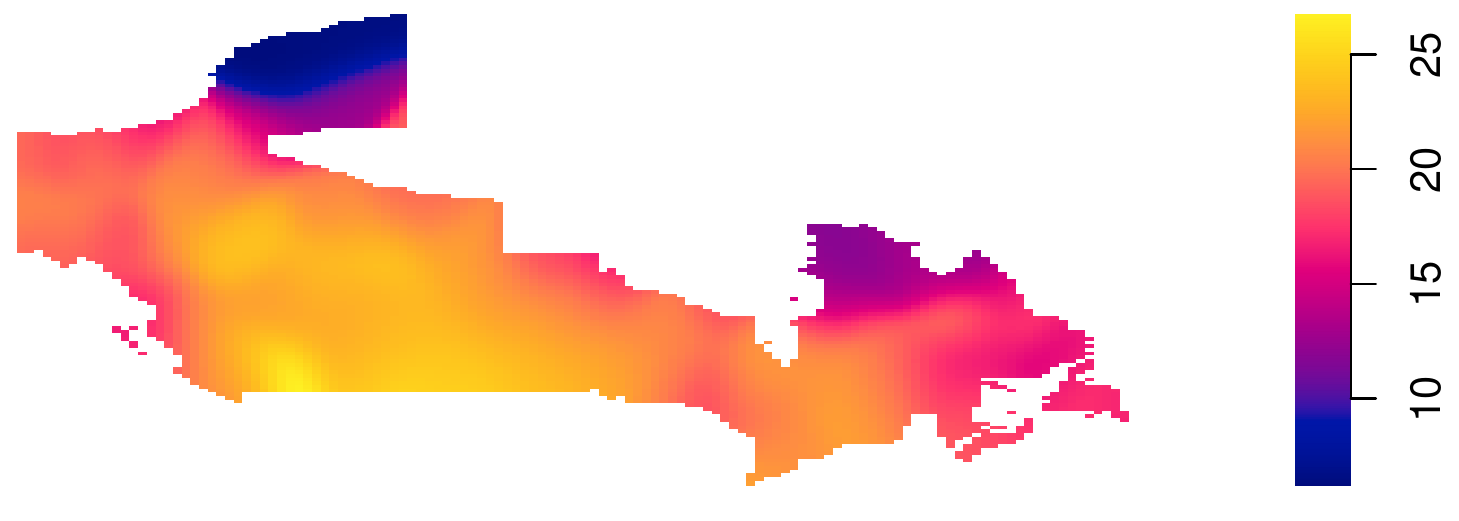}}}
	\hspace*{0.6cm}
	\subfigure[Mean noon-24 houre precipitation registered in June 2015 after a Gaussian smoothing with $\sigma=2$ (in millimetres).]{
		\resizebox*{.45\textwidth}{!}{\includegraphics[angle=90]{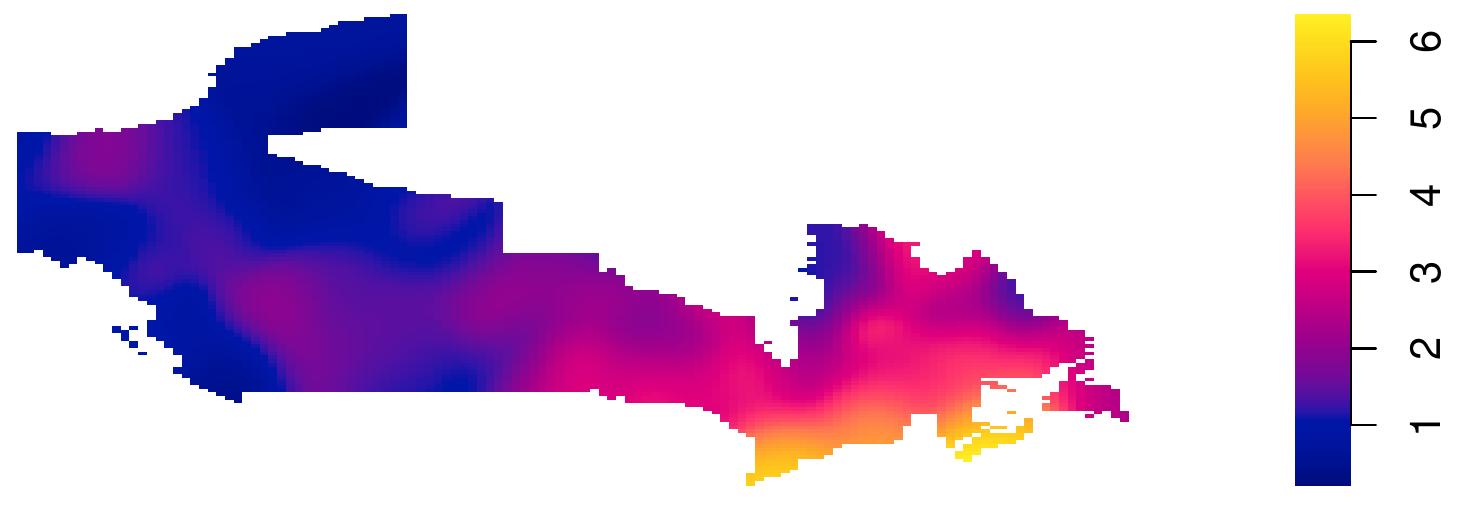}}}
	\caption{Covariates to be used in the intensity estimation of the wildfires in Canada during June 2015.}\label{fg:cov}
\end{figure}

We are interested in studying the spatial influence on some of these me\-teo\-ro\-lo\-gi\-cal variables on the distribution of wildfires. The wildfire data set and also a complete meteorological information from the last decades is available at the Canadian Wildland Fire Information System website (\url{http://cwfis.cfs.nrcan.gc.ca/home}). The fire season in Canada lasts from late April until August, with a peak of activity in June and July, hence we are interested in analysing the influence of meteorological covariates on wildfires during June 2015 (see Fig. \ref{fg:incjun2015}), and we focus our attention on temperature and precipitation (see Fig. \ref{fg:cov}). 

Our theoretical context \eqref{eq:int_rho}, is defined to use one single one-dimensional covariate in the model. As the temperature seems to have a higher influence on the distribution of the wildfires, we perform the estimation including it as the explanatory covariate; remark that instead of considering the maximum value during June in every point of the region, we have computed the third quartile in order not to deal with extreme values. We want to compare not only the results obtained using the different bandwidth selectors described in the previous sections but also, in order to detect the influence of the covariate, we have included the nonparametric kernel intensity estimation proposed in \cite{Diggle1985} that uses only the point pattern coordinates to compute the intensity estimation.

\begin{figure}[H]
	\begin{center}
		\subfigure[Diggle]{
			\resizebox*{.45\textwidth}{!}{
				\includegraphics[angle=90,scale=0.24]{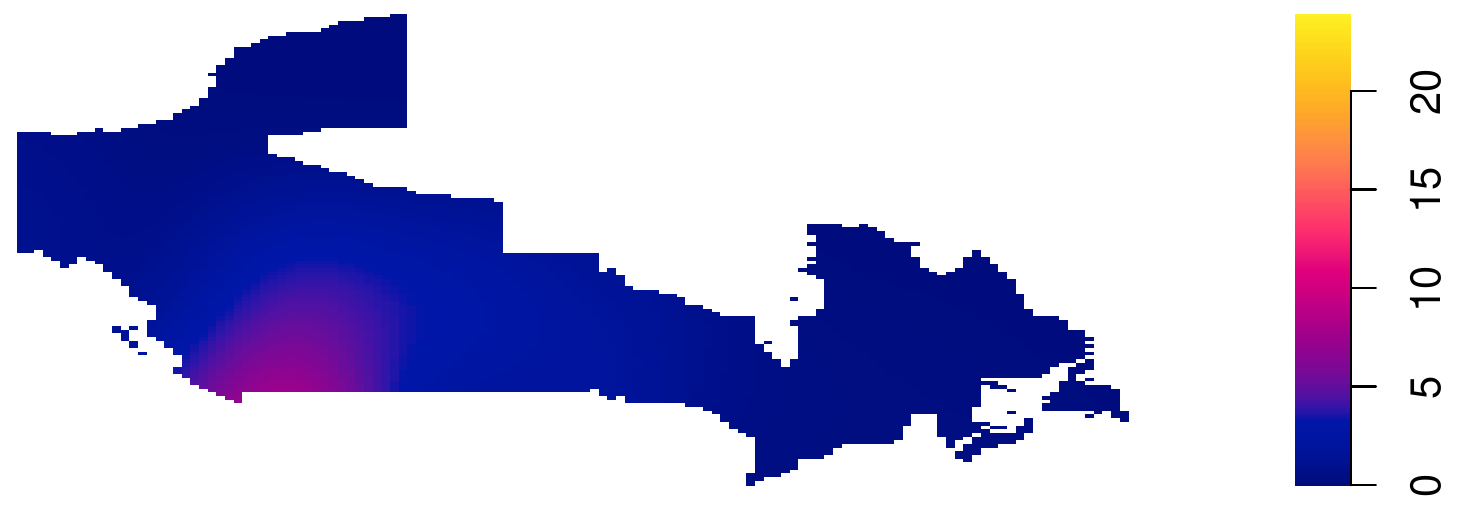}}}\hspace*{0.6cm}
		\subfigure[$\hat{h}_{\textup{Silv}}$]{
			\resizebox*{.45\textwidth}{!}{
				\includegraphics[angle=90,scale=0.24]{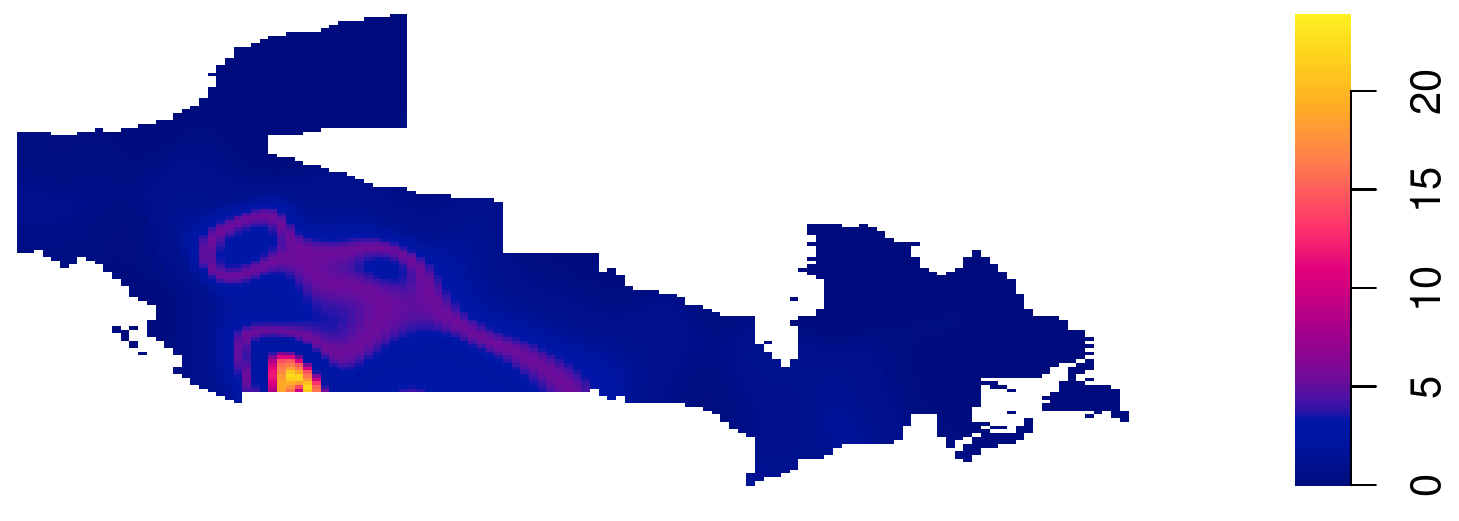}}}
		\vspace*{0.3cm}
		\subfigure[$\hat{h}_{\textup{RT}}$]{
			\resizebox*{.45\textwidth}{!}{
				\includegraphics[angle=90,scale=0.24]{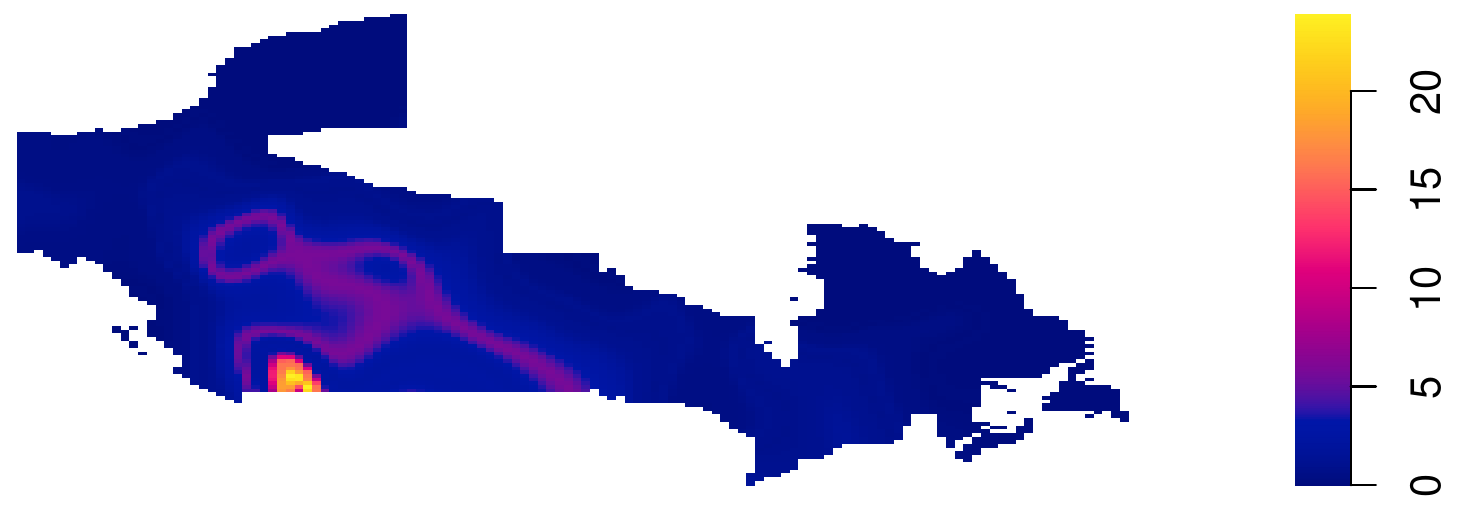}}}\hspace*{0.6cm}
		\subfigure[$\hat{h}_{\textup{Boot}}$]{
			\resizebox*{.45\textwidth}{!}{
				\includegraphics[angle=90,scale=0.24]{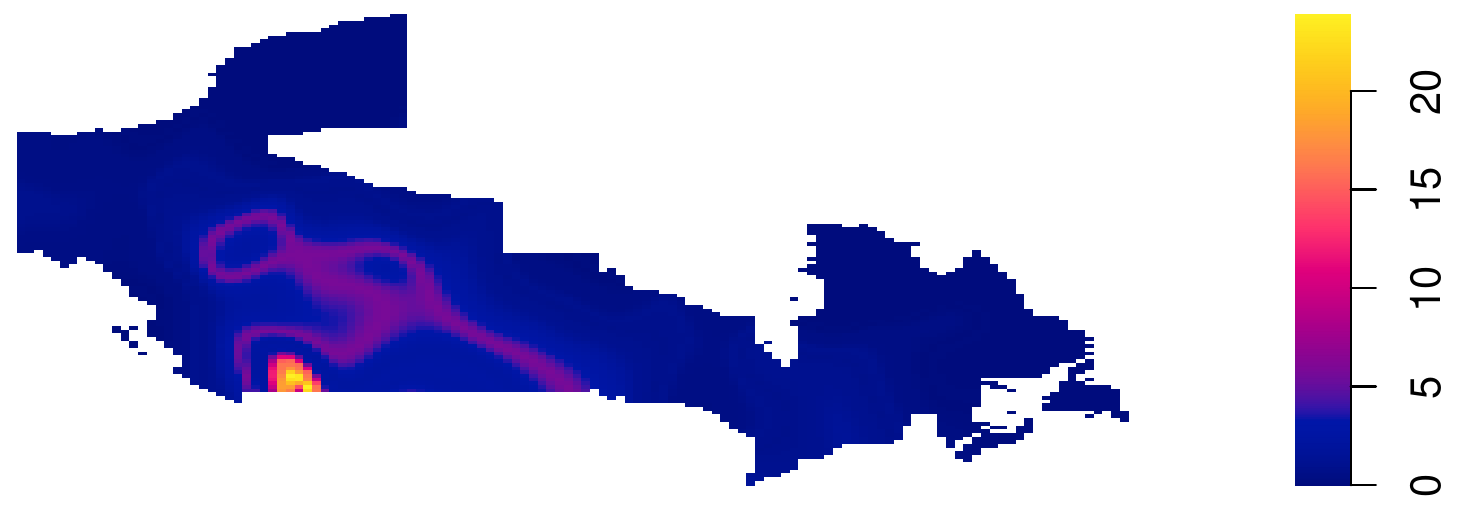}}}
		
		\caption{Diggle's estimation without covariate information (a), and estimations with the different bandwidth selectors using the temperature as the covariate ((b), (c) and (d)).}\label{fg:esttemp}
	\end{center}
\end{figure}

Regarding Fig. \ref{fg:incjun2015} and Fig. \ref{fg:esttemp} we can assure that the information given by the temperature is relevant for the estimation. Indeed, we can see in (a) that Diggle's proposal can barely identify the area with more fires, while when using this extra information, the estimation seems to be more suitable with the point process pattern. Among the three bandwidth selectors we cannot identify in this specific example one of them outperforming better than the others, actually the resulting estimates are very similar because the bandwidth values are close. 

In Section \ref{sc:ext} we detail some possible procedures to be able to use several covariates in this context. One possibility is to perform a Principal Component Analysis (PCA) for a set of covariates keeping the first principal component for the estimation. In this particular example of wildfire in Canada, the result is $\textup{ACP1}=0.991*\texttt{Temp} + 0.131*\texttt{Precip}$, which explains the 93\% of the total variance and the correlations with Temperature and Precipitation were, respectively, $0.9993$ and $0.4323$. This leads to the fact that temperature and the first principal component are almost equal, so we will obtain similar results.

In that same section related to multi-dimensional covariates, we also propose a multivariate extension of our model, where the use of several covariates is allowed. See the model description in \eqref{eq:int_multiv} and the multivariate intensity estimator in \eqref{eq:dens_est_multiv}. So, even if the asymptotic theory is not yet done for the multivariate context, we have developed the tools to be able to apply the multivariate extension in a practical context, including in this example temperature and precipitation all together in the estimation.

\begin{figure}[H]
	\centering
	\includegraphics[scale=0.45,angle=90]{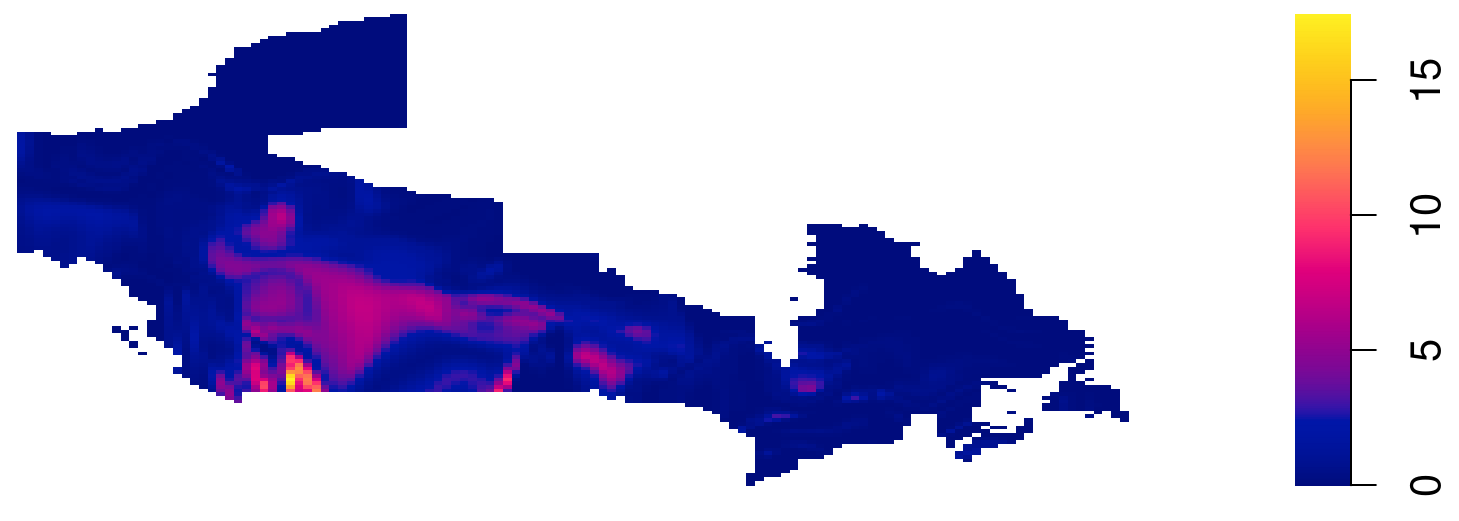}
	\caption{Intensity estimation performed with two covariates, temperature and precipitation.}\label{fg:estmultiv}
\end{figure}

We have to remark that we need a two-dimensional bandwidth matrix to be able to compute our multivariate estimator. To develop specific asymptotic theory supporting bandwidth selection in this multivariate context is beyond the scope of this paper. For this example we just consider the multivariate plug-in rule defined for density estimation in \cite{DuongHazelton} and implemented in the function \texttt{Hpi} of \cite{ks}. Note that his choice is not optimally designed for our estimator but it allows to complete our illustration.  

Fig. \ref{fg:estmultiv} shows the result of this multivariate estimation. Comparing it with Fig. \ref{fg:esttemp}, we can see that the resulting intensity seems to better represent the wildfire point pattern in Fig. \ref{fg:incjun2015}. This indicates that actually, both covariates seem to contribute significantly in the wildfire distribution.

\section{Further extensions}\label{sc:ext}
\subsection{Spatio-temporal point processes}
Spatio-temporal point processes are the most common generalisation of the spatial ones. To the extent of our knowledge spatio-temporal point processes have not been addressed so far for the covariate model we are focused on in this work, even though ths extension seems still natural. 

As a first approach let us define a spatio-temporal intensity depending on a covariate:
\[\lambda(u,t)= \rho(Z(u,t)),\]
where $u$ and $t$ are, respectively, the spatial and temporal coordinates, $Z: W\times T \subset \mathbb{R}^2\times\mathbb{R}\longrightarrow \mathbb{R}$ is the covariate and $\rho$ is an unknown real function. As it has been done before, we assume that $Z$ is known so, we only need to estimate $\rho$ in order to obtain an estimator of $\lambda$. 

As we have in this situation a covariate define in the spatio-temporal domain, the theory previously developed in this paper can be immediately applied to this new situation, reproducing all the results just taking into account the change of dimension in the domain of the covariate. 

Another possible framework in this spatio-temporal context can be the covariate having only a spatial or a temporal dependence, i.e, that $Z$ is either a function of space or either a function of time but not both together. Hence, the intensity function is 
\[\lambda(z,t)=\rho(Z(x),t)) \mbox{ or } \lambda(x,z)=\rho(x,Z(t)).\]
Remark that now $\rho$ is still a real function but with multivariate domain. In this case, following the literature about kernel intensity estimation in general spatio-temporal point processes, we propose the following estimators:
\[\hat{\rho}(z,t)=\sum_{i=1}^N \frac{1}{g^\star(Z_i)}K_h\left(z-Z_i\right)K_s(t-t_i) \mbox{ and }\] 
\[ \hat{\rho}(x,z)=\sum_{i=1}^N \frac{1}{g^\star(Z(t_i))}|H|^{-1/2}L\left(H^{-1/2}(x-X_i)\right)K_s(Z(t)-Z(t_i)),\]
where $(X_1,t_1),\ldots, (X_N,t_N)$ is the spatio-temporal pattern, $s$ is a univariate bandwidth parameter, $L$ is a radially symmetric bivariate density function, $H$ is a two-dimensional bandwidth matrix, and $Z_i$ and $g^\star$ are previously defined in the manuscript. The theoretical developments done in this work may be extended to these situations with the appropriate regularity conditions.

\subsection{Higher covariate dimension}
Although the framework we have set up in this paper is defined to use one single one-dimensional covariate, it is of interest to think about how this can be extended to the multivariate case. Hence, let us denote by $\mathbf{Z}=(Z_1, \ldots, Z_p)$ a $p$-dimensional covariate providing possible significant information about the process, where $Z_i:W\subset \mathbb{R}^2\rightarrow \mathbb{R}$ are one-dimensional continuous covariates, $\forall i \in \{1,\ldots,p\}$ fulfilling the same conditions as the initial $Z:W\subset \mathbb{R}^2\rightarrow \mathbb{R}$.

\subsubsection{Combinations of covariates}
A first approach to include the $p$ univariate covariates in the model, is to define a linear combination $a_1Z_1+\ldots+a_pZ_p$, where a procedure to estimate the coefficients needs to be determined, for example performing a principal component analysis (PCA) as we have illustrated in the previous section for our real data set. We choose the first component and we use its information in the estimation procedure, following our methodology previously presented in this paper. Note that the optimality criterion for selecting the best linear combination based on a direct PCA on the vector of covariates does not take into account its influence on the point process of interest, so other specific approaches should be explored to obtain better combinations.

\subsubsection{Multivariate framework}
Another possible solution to include several covariates is linked to the idea we have taken advantage from of the relationship between the density and the intensity functions. Hence, a multivariate version of \eqref{eq:dens_est} can be defined, where the use of several covariates would be allowed:
\begin{equation}\label{eq:int_multiv}
\lambda(u)=\rho(\mathbf{Z}(u)).
\end{equation}

Here we will briefly introduce the results to perform this extension, even though, as it is out of the scope of this paper, we are not developing everything in detail.

We need to introduce some notation generalising the one used in Section 3. The spatial cumulative distribution function in this multivariate context can be defined as

\begin{equation*}
G(z)=\frac{1}{|W|}\int_W{1_{\{\mathbf{Z}(u)\leq z\}}du},
\end{equation*}
where $\mathbf{Z}(u)\leq z$ refers to $(Z_1(u)\leq z_1)\cap \ldots \cap(Z_p(u)\leq z_p)$ with $z\in\mathbb{R}^p$. We still use $g$ to denote its first derivative, and $G^\star$ and $g^\star$ their non-normalised versions. 

Now we aim an extension of Theorem A.1 and Theorem A.2 in Appendix A to the multivariate case, for which we recall the results in \cite{DaleyVereJones1988} and \cite{Reiss2012}.

\begin{theorem}
	Let $X$ be a spatial point process in $W\subset \mathbb{R}^2$ with intensity function of the form \eqref{eq:int_rho} for some real-valued function $\rho$ and $\mathbf{Z}:W\subset \mathbb{R}^2 \rightarrow \mathbb{R}^p$ a continuous function. Then $\mathbf{Z}(X)$ is a $p$-dimensional point process in $\mathbb{R}^p$ with intensity function $\rho g^\star$. Moreover, if the original point process is Poisson, then the transformed one is also Poisson.
\end{theorem}

\begin{theorem}
	Let $W\subset \mathbb{R}^2$ be a bounded subset, $\mathbf{Z}:W\subset \mathbb{R}^2 \rightarrow \mathbb{R}^p$ a measurable, Lipschitz and differentiable function with non-zero Jacobian in every point of $W$, $J\mathbf{Z}(u)\neq0$. Then, for any integrable function $l:W\rightarrow\mathbb{R}$, in our particular case $l(u)=\lambda(u)(J \mathbf{Z}(u))^{-1}:W\rightarrow\mathbb{R}^+$, it holds:
	\[\int_W\lambda(u)du=\int_W\lambda(u)(J \mathbf{Z}(u))^{-1}J\mathbf{Z}(u)du=\int_{\mathbb{R}^p}\int_{\mathbf{Z}^{-1}(y)}\lambda(u)(J \mathbf{Z}(u))^{-1}dH^p(u)dy,\]
	where $\mathbf{Z}^{-1}(y)=\{u\in W / \mathbf{Z}(u)=y\}$ and $dH^p$ is the $p$-dimensional Hausdorff measure.
\end{theorem}

Applying this result to the non-normalised version of the spatial cumulative distribution function defined above, we have:
\begin{equation*}
G^{\star}(z)=\int_{-\infty}^{z_1}\cdots\int_{-\infty}^{z_p}\int_{\mathbf{Z}^{-1}(y)} (J\mathbf{Z}(u))^{-1}dH^p(u)dy_1,\ldots dy_p, 
\end{equation*}
and deriving with respect to $z$, we get $g^\star(z)=\int_{\mathbf{Z}^{-1}(z)}(J\mathbf{Z}(u))^{-1}dH^p(u)$.

We can now rewrite the relationship between the original spatial point process intensity and the transformed one through an integral, in a similar way we have done in the Appendix A for the univariate case, implying that the expected number of events in the corresponding region is the same in both processes:
\begin{align*}
& m:=\int_W{\lambda(u)du}=\int_{\mathbb{R}^p}\int_{\mathbf{Z}^{-1}(y)}\lambda(u)(J \mathbf{Z}(u))^{-1}dH^p(u)dy\\
&=\int_{\mathbb{R}^p}\int_{\mathbf{Z}^{-1}(y)}\rho(\mathbf{Z}(u))(J\mathbf{Z}(u))^{-1}dH^p(u)dy=\int_{\mathbb{R}^p}\rho(y)\int_{\mathbf{Z}^{-1}(y)}(J\mathbf{Z}(u))^{-1}dH^p(u)dy\\
&=\int_{\mathbb{R}^p}\rho(y)g^{\star}(y)dy.
\end{align*}

Once we have established the appropriate framework, we propose the following estimator for the associated ``artificial'' multivariate density, $f(\cdot)=\frac{\rho(\cdot)g^\star(\cdot)}{m}$:
\begin{equation}\label{eq:dens_est_multiv}
\hat{f}_{h}(z)=g^\star(z) \frac{1}{N}\sum_{i=1}^N\frac{1}{g^\star(Z_i)}\mathbf{K}_H\left(z-Z_i\right)1_{\{N\neq 0\}},
\end{equation}
and the extension of Baddeley's proposal to estimate $\rho$ can be derived as:
\begin{equation}\label{rhoestmultiv}
\hat{\rho}_H(z)=\sum_{i=1}^N\frac{1}{g^\star(Z_i)}\mathbf{K}_H(z-Z_i),
\end{equation}
where now $\mathbf{K}$ should be a multivariate radially symmetric kernel function and $H$ a $p$-dimensional bandwidth matrix.

The analogous theoretical developments presented in Section 3 and Section 4 can be extended to the multivariate situation following the steps we have already detailed and also applying some specifically designed statistical tools for the multivariate analysis used in \cite{Scott1992}, \cite{CucalaThesis} and \cite{Isa2015}.

\section{Conclusions}
We have considered kernel intensity estimation in the context of spatial point processes with covariates. We have set up an innovative theoretical framework that has allowed us to detail the expressions of the MSE, MISE and AMISE for our intensity estimator. Furthermore we have proposed a consistent smooth bootstrap procedure, and two new data-driven bandwidth selection methods. We have also studied their behaviour and compare them with the previous methods used in this context through an extensive simulation study; the overall conclusion being a better performance of the bootstrap-based bandwidth. The application to a real data set also shows that our proposals are competitive with the existing ones and even better than kernel intensity estimators based only on information provided by the locations of the events. Also in terms of computational cost our proposals are faster or competitive with the existing ones. Finally we propose some extensions of our methodology for spatio-temporal point processes and in the field of multivariate analysis including several covariates in the model.

\section{Acknowledgments}
The authors acknowledge the support from the Spanish Ministry of Economy and Competitiveness, through grants number MTM2013-41383P and MTM2016-76969P, which includes support from the European Regional Development Fund (ERDF). Support from the IAP network StUDyS from Belgian Science Policy (P6/07), is also acknowledged. M.I. Borrajo has been supported by FPU grant (FPU2013/00473) from the Spanish Ministry of Education. The authors also acknowledge the Canadian Wildland Fire Information System for their activity in recording and freely providing all the real data used in this paper.

\bibliographystyle{apalike}

\newpage
\appendix
\renewcommand{\thesection}{Appendix \Alph{section} -}
\newtheorem{theorem2}{Theorem A.}
\section{Useful tools for spatial processes}
Following \cite{DaleyVereJones1988} and \cite{Reiss2012} we develop the following result that allows us to establish the model for the transformed point processes in the covariate space:
\begin{theorem2}
	Let $X$ be a spatial point process in $W\subset \mathbb{R}^2$ with intensity function of the form \eqref{eq:int_rho} for some real-valued function $\rho$. Then, $Z(X)$ is a univariate point process in $\mathbb{R}$ with intensity function $\rho g^\star$.
	Moreover, if the original point process is Poisson, then the transformed one preserve this property and it is also Poisson.
\end{theorem2}

The relationship between $X$ and $Z(X)$ can be extended to the expected number of events trough this result adapted from \cite{Federer1969}:
\begin{theorem2}\label{th:coarea}
	Let $W\subset \mathbb{R}^2$ be a bounded subset, $Z:W\subset \mathbb{R}^2 \rightarrow \mathbb{R}$ a differentiable function with non-zero gradient in every point of $W$. Then, for any integrable function $l:W\rightarrow\mathbb{R}$, in our particular case $l(u)=\lambda(u)||\nabla Z(u)||^{-1}:W\rightarrow\mathbb{R}^+$, it holds:
	\[\int_W\lambda(u)du=\int_W\lambda(u)||\nabla Z(u)||^{-1}||\nabla Z(u)||du=\int_{\mathbb{R}}\int_{Z^{-1}(y)}\lambda(u)||\nabla Z(u)||^{-1}dH(u)dy,\]
	where $Z^{-1}(y)=\{u\in W / Z(u)=y\}$ and $dH$ is the one-dimensional Hausdorff measure.
\end{theorem2}

Applying this result to the non-normalised spatial cumulative distribution function $G^\star(z)=|W|G(z)$ we have:
\begin{align*}
G^{\star}(z)&=\int_W{1_{\{Z(u)\leq z\}}du}=\int_W{1_{\{Z(u)\leq z\}}||\nabla Z(u)||^{-1}||\nabla Z(u)||du}\\
&=\int_{\mathbb{R}}{\int_{Z^{-1}(y)}1_{\{Z(u)\leq z\}}||\nabla Z(u)||^{-1}dH(u)}dy=\int_{-\infty}^{z}\int_{Z^{-1}(y)}||\nabla Z(u)||^{-1}dH(u)dy.
\end{align*}
Deriving with respect to $z$, we get
$\displaystyle g^{\star}(z):=(G^{\star})^{'}(z)=\int_{Z^{-1}(z)}||\nabla Z(u)||^{-1}dH(u)$.\\

We can now rewrite the relationship between the original spatial point process intensity and the transformed one through an integral, which also implies that the expected number of events in the corresponding region is the same in both processes:
\begin{align*}
& m:=\int_W{\lambda(u)du}=\int_{\mathbb{R}}\int_{Z^{-1}(y)}\lambda(u)||\nabla Z(u)||^{-1}dH(u)dy=\int_{\mathbb{R}}\int_{Z^{-1}(y)}\rho(Z(u))||\nabla Z(u)||^{-1}dH(u)dy\\
&=\int_{\mathbb{R}}\rho(y)\int_{Z^{-1}(y)}||\nabla Z(u)||^{-1}dH(u)dy=\int_{\mathbb{R}}\rho(y)g^{\star}(y)dy.
\end{align*}
Hence, we can finally have a univariate nonhomogeneous \textbf{Poisson} point process.

\renewcommand{\thesection}{Appendix \Alph{section} -}
\section{Proof of Theorem \ref{th:mse}} \label{ap2}

Recall that $N$ is not a constant but a random variable, so we have a double stochastic scenario, on one side the randomness provided by $N$ and on the other the randomness of the point process. To deal with this we use the conditional mean and we consider also some tools related to real number series.

Consider first the mean,
\begin{align*}
E\left[\hat{f}_h(z)\right]&=E\left[E\left[\hat{f}_h(z)|N=n>0\right]\right]=\sum_{n=1}^\infty E\left[\hat{f}_h(z)|N=n\right]\mathbb{P}\left(N=n\right)\nonumber\\
&=\sum_{n=1}^\infty E\left[\hat{f}_h(z)|N=n\right]\frac{e^{-m}m^n}{n!}=\sum_{n=1}^\infty \frac{g^\star(z)}{m}\left(K_h\circ\rho\right)(z)\frac{e^{-m}m^n}{n!}\nonumber \\
&=\frac{g^\star(z)(K_h \circ \rho)(z)}{m}e^{-m}\left(\sum_{n=1}^\infty\frac{m^n}{n!}-1\right)=\frac{g^\star(z)(K_h \circ \rho)(z)}{m}\left(1-e^{-m}\right),
\end{align*}

where we have used that
\begin{align*}
&E\left[\hat{f}_h(z)|N=n\right]=E\left[g^\star(z)\frac{1}{n}\sum_{i=1}^n\frac{1}{g^\star(Z_i)}K_h(z-Z_i)\right]=g^\star(z)E\left[\frac{1}{g^\star(Z_1)}K_h(z-Z_1)\right]\\
&=g^\star(z)\int_{\mathbb{R}}\frac{1}{g^\star(s)}K_h(z-s)\frac{\rho(s)g^\star(s)}{m}ds=\frac{g^\star(z)}{m}\left(K_h\circ\rho\right)(z).
\end{align*}

Next, the variance:
\begin{align*}
& Var\left[\hat{f}_h(z)\right]=E\left[\hat{f}^2_h(z)\right]-E\left[\hat{f}_h(z)\right]^2\nonumber=A(m)\frac{(g^\star(z))^2}{n}\left(K^2_h \circ \frac{\rho}{g^\star}\right)(z)-\nonumber \\
&-\left(A(m)+e^{-m}-1\right)\frac{(g^\star(z))^2}{m^2}\left(K_h\circ \rho\right)^2(z)-\frac{(g^\star(z))^2\left(K_h\circ \rho\right)^2(z)}{m^2}(1-e^{-m})^2\nonumber \\
&=A(m)\frac{(g^\star(z))^2}{n}\left(K^2_h \circ \frac{\rho}{g^\star}\right)(z)-(A(m)+e^{-2m}-e^{-m})(g^\star(z))^2(K_h\circ \rho)^2(z)
\end{align*}

where we have used the notation $A(m):=\left[\frac{1}{N}1_{\{N\neq 0\}}\right]$ and,
\begin{align*}
&E\left[\hat{f}^2_h(z)\right]=E\left[E\left[\hat{f}_h^2(z)|N=n>0\right]\right]=\sum_{n=1}^\infty E\left[\hat{f}_h^2(z)|N=n\right]\mathbb{P}\left(N=n\right)\\&
=\sum_{n=1}^\infty E\left[\hat{f}_h^2(z)|N=n\right]\frac{e^{-m}m^n}{n!}=\frac{(g^\star(z))^2}{m}\left(K^2_h \circ \frac{\rho}{g^\star}\right)(z)\sum_{n=1}^\infty\frac{e^{-m}m^n}{n n!}+\\
&+\frac{(g^\star(z))^2}{m^2}\left(K_h\circ \rho\right)^2(z)\sum_{n=1}^\infty\frac{n-1}{n}\frac{e^{-m}m^n}{n!}=\frac{(g^\star(z))^2}{m}\left(K^2_h \circ \frac{\rho}{g^\star}\right)(z)E\left[\frac{1}{N}1_{\{N\neq 0\}}\right]+\\
&+\frac{(g^\star(z))^2}{m^2}\left(K_h\circ \rho\right)^2(z)E\left[\frac{N-1}{N}1_{\{N\neq 0\}}\right]=A(m)\frac{(g^\star(z))^2}{m}\left(K^2_h \circ \frac{\rho}{g^\star}\right)(z) - \\
&- \left(A(m)+e^{-m}-1\right)\frac{(g^\star(z))^2}{m^2}\left(K_h\circ \rho\right)^2(z),
\end{align*}

as well as

\begin{align*}
&E\left[\hat{f}^2_h(z)|N=n\right]=E\left[\left(g^\star(z)\right)^2\frac{1}{n^2}\left(\sum_{i=1}^n\frac{1}{g^\star(Z_i)}K_h(z-Z_i)\right)^2\right]\\
&=\frac{(g^\star(z))^2}{n^2}E\left[\left(\sum_{i=1}^n\frac{1}{g^\star(Z_i)}K_h(z-Z_i)\right)^2\right]\\
&=\frac{(g^\star(z))^2}{n^2}E\left[\sum_{i=1}^n\frac{1}{(g^\star(Z_i))^2}K^2_h(z-Z_i)+\sum_{i\neq j}\frac{1}{g^\star(Z_i)}\frac{1}{g^\star(Z_j)}K_h(z-Z_i)K_h(z-Z_j)\right]\\
&=\frac{(g^\star(z))^2}{n}E\left[\frac{1}{(g^\star(Z_1))^2}K_h^2(z-Z_1)\right]+\frac{(g^\star(z))^2}{n^2}n(n-1)\left(E\left[\frac{1}{g^\star(Z_1)}K_h(z-Z_1)\right]\right)^2\\
&=\frac{(g^\star(z))^2}{n}\int_{\mathbb{R}}\frac{1}{(g^\star(s))^2}K^2_h(z-s)\frac{\rho(s)g^\star(s)}{m}ds+\frac{(g^\star(z))^2(n-1)}{n}\left(\int_{\mathbb{R}}\frac{1}{g^\star(s)}K_h(z-s)\frac{\rho(s)g^\star(s)}{m}ds\right)^2\\
&=\frac{(g^\star(z))^2}{mn}\left(K^2_h \circ \frac{\rho}{g^\star}\right)(z)+\frac{(g^\star(z))^2(n-1)}{nm^2}\left(K_h\circ \rho\right)^2(z).
\end{align*}

Lastly compute the MSE in terms of bias and variance. We apply a Taylor expansion of second order to the mean and of first order to the variance, getting:
\begin{align*}
& E\left[\hat{f}_h(z)\right]=\frac{g^\star(z)}{m}\left(K_h\circ \rho\right)(z)(1-e^{-m})=\frac{g^\star(z)}{m}\left(\rho(z)+\frac{1}{2}h^2\mu_2(K)\rho^{''}(z)+o(h^2)\right)(1-e^{-m})\\
& Bias \left[\hat{f}_h(z)\right]=\frac{g^\star(z)}{m}\left(-e^{-m}\rho(z)+(1-e^{-m})\frac{1}{2}h^2\rho^{''}(z)\mu_2(K)+o(h^2(1-e^{-m}))\right)\\
& Var\left[\hat{f}_h(z)\right]=A(m)\frac{(g^\star(z))^2}{m}\left(K^2_h \circ \frac{\rho}{g^\star}\right)(z)-(A(m)+e^{-2m}-e^{-m})(g^\star(z))^2(K_h\circ \rho)^2(z)\\
&\hspace*{2.15cm}=A(m)\frac{(g^\star(z))^2}{m}\frac{1}{h}\frac{\rho}{g^\star}(z)R(K)+o\left(\frac{A(m)}{mh}\right)
\end{align*}
where $\mu_2(K)=\int_{\mathbb{R}}z^2K(z)dz$ and $R(K)=\int_{\mathbb{R}}K^2(u)du$. Hence,
\begin{align*}
MSE\left[\hat{f}_h(z)\right]&=e^{-2m}f^2(z)+(1-e^{-m})^2\frac{h^4}{4}\left(\frac{\rho^{''}(z)g^\star(z)}{m}\right)^2\mu_2^2(K)-\nonumber\\
&-e^{-m}(1-e^{-m})h^2\mu_2(K)\frac{(g^\star(z))^2\rho(z)\rho^{''}(z)}{m^2}+\frac{A(m)}{h}f(z)R(K)+\nonumber \\
&+o(h^2(1-e^{-m})e^{-m})+o(h^4(1-e^{-m})^2)+o\left(\frac{A(m)}{mh}\right);
\end{align*}

from which we can easily derive its asymptotic version
\begin{align*}
AMSE\left[\hat{f}_h(z)\right]&=e^{-2m}f^2(z)+(1-e^{-m})^2\frac{h^4}{4}\left(\frac{\rho^{''}(z)g^\star(z)}{m}\right)^2\mu_2^2(K)-\nonumber\\
&-e^{-m}(1-e^{-m})h^2\mu_2(K)\frac{(g^\star(z))^2\rho(z)\rho^{''}(z)}{m^2}+\frac{A(m)}{h}f(z)R(K).
\end{align*}

\section{Proof of Theorem \ref{th:mseboot}}\label{ap3}
To compute the $MSE^\ast$ of \eqref{eq:dens_est_boot} we follow the same structure we have established for Theo\-rem \ref{th:mse}, with the difference that we are now under the bootstrap distribution. 

The pointwise mean is:
\begin{align*}
E^\ast\left[\hat{f}^\ast_h(z)\right]&=E^\ast\left[E^\ast\left[\hat{f}^\ast_h(z)|N^\ast=n^\ast>0\right]\right]=\sum_{n^\ast=1}^\infty E^\ast\left[\hat{f}_h^\ast(z)|n^\ast=n^\ast\right]\mathbb{P}^\ast\left(N^\ast=n^\ast\right)\\
&=\sum_{n^\ast=1}^\infty E^\ast\left[\hat{f}_h^\ast(z)|N^\ast=n^\ast\right]\frac{e^{-\hat{m}}\hat{m}^{n^\ast}}{n^\ast!}=\sum_{n^\ast=1}^\infty\frac{g^\star(z)}{\hat{m}}(K_h\circ \hat{\rho}_b)(z)\frac{e^{-\hat{m}}\hat{m}^{n^\ast}}{n^\ast!}\\
&=\frac{g^\star(z)}{\hat{m}}(K_h\circ \hat{\rho}_b)(z)(1-e^{-\hat{m}}), \mbox{ where we have used that}
\end{align*}

\begin{align*}
&E^\ast\left[\hat{f}^\ast_h(z)|N^\ast=n^\ast>0\right]=E^\ast\left[g^\star(z)\frac{1}{n^\ast}\sum_{i=1}^{n^\ast}\frac{1}{g^\star(Z_i^\ast)}K_h(z-Z_i^\ast)\right]=g^\star(z)E^\ast\left[\frac{1}{g^\star(Z_1^\ast)}K_h(z-Z_1^\ast)\right]\\
&=\frac{g^\star(z)}{\hat{m}}\int{K_h(z-s)\hat{\rho}_b(s)ds}=\frac{g^\star(z)(K_h\circ\hat{\rho}_b)(z)}{\hat{m}}.
\end{align*}

We now apply a second order Taylor expansion to the mean in order to obtain the bias:
\begin{align*}
Bias^\ast\left[\hat{f}^\ast_h(z)\right]=-e^{-\hat{m}}\frac{\hat{\rho}_b(z)g^\star(z)}{\hat{m}}+\frac{h^2}{2}\frac{\hat{\rho}_b^{''}(z)g^\star(z)}{\hat{m}}\mu_2(K)(1-e^{-\hat{m}})+o(h^2(1-e^{-\hat{m}})).
\end{align*}

To derive the variance we work as follows:
\begin{align*}
Var^\ast\left[\hat{f}_h^\ast(z)\right]&=E^\ast\left[\hat{f}_h^{\ast \: 2}(z)\right]-E^{\ast}\left[\hat{f}_h^\ast(z)\right]^2\\
&=\frac{(g^\star(z))^2}{\hat{m}}\left(K_h\circ\frac{\hat{\rho}_b}{g^\star}\right)(z)A(\hat{m})-\frac{(g^\star(z))^2}{\hat{m}^2}(K_h\circ\hat{\rho}_b)^2(z)(A(\hat{m})+e^{-2\hat{m}}-e^{-\hat{m}}).
\end{align*}

Applying a first order Taylor expansion we get:
\begin{align*}
Var^\ast\left[\hat{f}_h^\ast(z)\right]=\tilde{f}_b(z)R(K)\frac{A(\hat{m})}{h}+o\left(\frac{A(\hat{m})}{\hat{m}h}\right),
\end{align*}
where we have used:
\begin{align*}
&E^\ast\left[(\hat{f}_h^\ast(z))^2\right]=E^\ast\left[E^\ast\left[\hat{f}_h^{\ast\:2}(z)|N^\ast=n^\ast\right]\right]=\sum_{n^\ast=1}^\infty E^\ast\left[\hat{f}_h^{\ast\:2}(z)|n^\ast=n^\ast\right]\mathbb{P}^\ast\left(N^\ast=n^\ast\right)\\
&=\sum_{n^\ast=1}^\infty E^\ast\left[\hat{f}_h^{\ast\:2}(z)|N^\ast=n^\ast\right]\frac{e^{-\hat{m}}\hat{m}^{n^\ast}}{n^\ast!}=\frac{(g^\star(z))^2}{\hat{m}}\left(K_h^2\circ \frac{\hat{\rho}_b}{g^\star}\right)(z)\sum_{n^\star=1}^\infty\frac{e^{-\hat{m}}\hat{m}^{n^\ast}}{n^\ast n^\ast!}+ \\
&+\frac{(g^\star(z))^2}{\hat{m}^2}(K_h\circ\hat{\rho}_b)^2(z)\sum_{n^\ast=1}^\infty\frac{n^\ast-1}{n^\ast}\frac{e^{-\hat{m}}\hat{m}^{n^\ast}}{n^\ast!}=\frac{(g^\star(z))^2}{\hat{m}}\left(K_h^2\circ \frac{\hat{\rho}_b}{g^\star}\right)(z)A(\hat{m})-\\
&-\frac{(g^\star(z))^2}{\hat{m}^2}(K_h\circ\hat{\rho}_b)^2(z)(A(\hat{m}+e^{-\hat{m}}-1), \mbox{ where}
\end{align*}
\begin{align*}
&E^\ast\left[\hat{f}^{\ast\:2}_h(z)|N^\ast=n^\ast\right]=E^\ast\left[\left(g^\ast(z)\frac{1}{n^\ast}\sum_{i=1}^{n^\ast}\frac{1}{g^\star(Z_i^\ast)}K_h(z-Z_i^\ast)\right)^2\right]\\
&=\frac{(g^\star(z))^2}{n^{\ast\:2}}E^\ast\left[\sum_{i=1}^{n^\ast}\frac{1}{(g^\star(Z_i^\ast))^2}K^2_h(z-Z_i^\ast)+\sum_{i\neq j}\frac{1}{g^\star(Z_i^\ast)}\frac{1}{g^\star(Z_j^\ast)}K_h(z-Z_i^\ast)K_h(z-Z_j^\ast)\right]\\
&=\frac{(g^\star(z))^2}{n^{\ast\:2}}\left(n^\ast E^\ast\left[\frac{1}{(g^\star(Z_1^\ast))^2}K_h^2(z-Z_1^\ast)\right]+n^\ast(n^\ast-1)E^\ast\left[\frac{1}{g^\star(Z_1^\ast)}K_h(z-Z_1^\ast)\right]^2\right)\\
&=\frac{(g^\star(z))^2}{n^{\ast}\hat{m}}\left(K_h^2\circ\frac{\hat{\rho}_b}{g^\star}\right)(z) + \frac{(g^\star(z))^2(n^\ast-1)}{n^\ast \hat{m}^2}(K_h\circ\hat{\rho}_b)^2(z).
\end{align*}

Finally, gathering square bias and variance together we obtain the $MSE^\ast$ in \eqref{eq:mseboot}.

\end{document}